\documentclass{aastex63}
\usepackage{natbib}
\begin{document}








\author{G. Gronoff}
\affil{Science Systems and Application Inc. Hampton, Va, USA}
\affil{Nasa Langley Research Center, Hampton, Va, USA}
\email{Guillaume.P.Gronoff@nasa.gov}

\author{R. Maggiolo}
\affil{Royal Belgian Institute for Space Aeronomy (BIRA-IASB), Brussels, Belgium.}

\author{G. Cessateur}
\affil{Royal Belgian Institute for Space Aeronomy (BIRA-IASB), Brussels, Belgium.}

\author{W.B. Moore}
\affil{Hampton University, Hampton, Virginia, USA}

\author{V. Airapetian}
\affil{American University, MD, USA}
\affil{Nasa GSFC/SEEC, Greenbelt, MD, USA}
\author{J. De Keyser}
\affil{Royal Belgian Institute for Space Aeronomy (BIRA-IASB), Brussels, Belgium.}
\affil{Centre for mathematical Plasma Astrophysics, Heverlee, Belgium}
\author{F. Dhooghe}
\affil{Royal Belgian Institute for Space Aeronomy (BIRA-IASB), Brussels, Belgium.}
\author{A. Gibbons}
\affil{Royal Belgian Institute for Space Aeronomy (BIRA-IASB), Brussels, Belgium. \\Universit\'e Libre de Belgique, Brussels, Belgium.}
\author{H. Gunell}
\affil{Royal Belgian Institute for Space Aeronomy (BIRA-IASB), Brussels, Belgium.}

\author{C.J. Mertens}
\affil{Nasa Langley Research Center, Hampton, Va, USA}

\author{M. Rubin}
\affil{Physikalisches Institut, University of Bern, Bern, Switzerland}

\author{S. Hosseini}
\affil{Jet Propulsion Laboratory, Pasadena, Ca, USA}

\shorttitle{GCR on Comet 67P}
\shortauthors{Gronoff et al.}



\title{The Effect of Cosmic Rays on Cometary Nuclei: I Dose deposition}

-



%
%
\begin{abstract}
Comets are small bodies thought to contain the most pristine material in the solar system. However, 
since their formation $\approx$4.5~Gy ago, they have been altered by different processes. While not exposed to much electromagnetic radiation, they experience intense particle radiation. Galactic cosmic rays and solar energetic particleshave a broad spectrum of energies and interact with the cometary surface and subsurface; they are the main source of space weathering for a comet in the Kuiper Belt or in the Oort cloud; and also affect the ice prior to the comet agglomeration.   While low energy particles interact only with the cometary surface, the most energetic ones deposit a significant amount of energy down to tens of meters. This interaction can modify the isotopic ratios in cometary ices and create secondary compounds through radiolysis, such as O$_2$ and H$_2$O$_2$ (paper II: Maggiolo et al., 2019). 
In this paper, we model the energy deposition of energetic particles as a function of depth using a Geant4 application modified to account for the isotope creation process. We quantify the energy deposited in cometary nucleus by galactic cosmic rays and solar energetic particles. The consequences of the energy deposition on the isotopic and chemical composition of cometary ices and their implication on the interpretation of cometary observations, notably of 67P/Churyumov Gerasimenko by the ESA/Rosetta spacecraft, will be discussed in Paper II.

\end{abstract}


\keywords{comets: general --- comets: individual(67P) --- Cosmic Rays}



\section{Introduction}
%
%
%
Comets formed in the early stage of the solar system and, today, reside in two mains reservoirs, the Kuiper Belt/Scattered disk (KB, 355 au) and the Oort Cloud (OC, 2000-200000 au). Orbital perturbations (for instance a collision or the gravitational force exerted by Jupiter) bring some comets closer to the Sun. If close enough to the Sun, their nuclei is heated by sunlight and cometary ice sublimates into space, while refractory grains are set free. 
It is during this stage that spacecraft can sample in-situ cometary material as the ESA/Rosetta probe did from August 2014 to September 2016 around comet 67P Churyumov/Gerasimenko (referred to as 67P in the following).
 67P is a  KB comet that has been gravitationally displaced into the inner Solar System by a close encounter with Jupiter in 1959 \citep{Maquet2015}. It is now orbiting with a semi-major axis of 3.46~au \citep{1985ltes.book.....C}.
The comet is estimated to lose from a few up to tens of meters of surface material during each orbit \citep{Bertaux2015,Keller2015}. It should thus have lost several tens of meters on average since insertion in its current orbit, with locally higher erosion rates, depending on the position on the surface of the nucleus \citep{Groussin2015}. The loss may actually be higher  since the comet was already outgassing when it reached the Jupiter region during its pre-1959 orbit \citep{matonti_bilobate_2019}.\\

Consequently, the neutral gas measured in-situ in the coma of comet 67P by the ROSINA experiment (and also MIRO, VIRITS, and Alice) onboard Rosetta likely originated several tens of meters beneath the primordial surface of the comet.
ROSINA observations provided evidence that this comet is formed from  pristine material which hasn't been significantly altered after its formation in the first Myr of the solar nebula stage. The high abundance of super-volatiles like CO and CO$_2$ \citep{LeRoy2015}, the detection of argon \citep{Balsiger2015}, of molecular nitrogen \citep{Rubin2015}, of molecular oxygen \citep{Bieler2015}, of a high D/H in HDO/H2O and D2O/HDO and HDS/H2S \citep{Altwegg2015,Altwegg2017}, and of hydrogen halides \citep{DhoogheEtAl2017,DeKeyserEtAl2017}, 
 coupled with the low density, high porosity and homogeneity of the nucleus \citep{Paetzold2016} and the absence of signatures of aqueous alteration \citep[see][]{Davidsson2016,bardyn2017carbon,Capaccioniaaa0628,QUIRICO201632} all indicate that comet 67P formed at low temperature and did not experience any substantial global scale heating after its formation.
This suggests that 67P is representative of the solar nebula material from which the solar system had formed. This has strong implications not only for how the measurements made in cometary environments can be used to constrain the proto-solar environment but also for the contribution of comets to Earth's composition.
For instance, the measurement of the D/H isotopic ratio in 67P \citep{Altwegg2015} suggests that comets cannot be considered as the main source of water on Earth. The discovery of significant amounts of O$_2$ in comets \citep{Bieler2015} was not predicted by astrochemical models and challenges our understanding of the chemistry of molecular clouds and of the proto-solar nebula. 
However, the Jupiter family comets (JFC, which include 67P) are a diverse groups. Indeed, even if Giotto measurements indicate that comet 1P/Halley contains similar amounts of O$_2$ \citep{Rubin2015}, different D/H ratios (lower than observed for 67P and compatible with the D/H ratio in the Earth's oceans) have been measured for other JFC comets like Hartley 2 \citep{Balsiger2015} and 46P/Wirtanen \citep{Dariusz2019}.
The causes for this diversity may be already present at the formation of these comets or may result from a different evolution after their formation.

Indeed, there is evidence that some processes alter cometary material after their formation \citep[e.g.][and references therein] {Stern2003, Guilbert-Lepoutre2015} without inducing a significant large scale heating of the nucleus. The main reservoirs of comets, the KB and OC, are sufficiently far from the Sun so that comets receive little heat. In addition, their small size (a few tens of kilometers or less) prevents comets from  transformed by internal (radiogenic) heating \citep{MousisEtAl2017}. However, the thermal environment of comets may have varied under the effect of passing stars and of supernova explosions in the vicinity of the solar system. \cite{Stern1988Natur} consider as very likely that OC comets have been heated at least once  by a luminous O star passing in the vicinity ($<$ 5 pc) of the solar system. This can raise the temperature of the OC from 5-6 K to $\sim$16 K during a period of the order of  $\sim3\times10^4$ years. For KB comets the estimated temperature increase is lower ($\sim$1 K) as their equilibrium temperature is higher (30-60 K) due to their location at shorter heliocentric distance. OC comets may also have experienced several but short-term ($\sim$0.1 year) temperature increases of a few tens of Kelvin due to nearby supernovae \citep{Stern1988Natur}. Due to the short duration of supernovae explosions, this type of heating would mostly affect the first meter below the surface of a cometary nucleus while the heat from a passing O star can penetrate down to a few tens of meters owing to its longer duration. Heating by passing stars and supernovae in the OC regions may have removed some part of the most volatile elements from their surface/subsurface.\\
While OC comets may have experienced collisions during their ejection from the inner young solar system to the OC \citep{Stern2001}, the OC is essentially a collisionless environment \citep{Stern1988Icar} due to its low density and to the low orbital velocity in this region \cite [see][and references therein]{Stern2003}. It is thus likely that OC comets haven't been significantly altered by collisions. \\
KB comets may have experienced more frequent collisions after they formed \citep[e.g.][]{Duncan2004} as both the density and orbital velocities are higher in this region which should result in a higher collision rate than in the OC.  
 The collisional history of of KB comets is still debated. Several scenarios  have been proposed to explain the formation of bilobate shaped comets like 67P. 
\cite{Davidsson2016} argues that Rosetta observations of comet 67P are in better agreement with a comet formation by agglomeration of primordial rubble-piles that remained in the protosolar nebulae after trans-Neptunian objects formation. This process should be slow enough, between $\sim$2.2 and 7.7 Myr after the incorporation of Ca-AL-rich inclusions from the protosolar nebula in 67P, in order to avoid thermal heating by short lived radionuclides \citep{Mousis2017}. Observations by the OSIRIS camera on-board Rosetta evidenced that 67P consists of two different objects that have formed a contact binary \citep{Massironi2015}. Bilobate comets are relatively common and \cite{Davidsson2016} suggested that the final stage of comet formation results in merging between lobes and that comets are unlikely to experience destructive collisions. 
Other authors come to the opposite conclusion: \cite{Morbidelli2015} and \cite{Rickman2015} estimated the probability that 67P survived collisions to 10$^{-4}$  if formed early in the protosolar nebula. \citet{JutziBenz2017} suggest that bilobate comets like 67P may have  formed due to sub-catastrophic collisions occurring later, which wouldn't lead to an alteration of the nucleus  in agreement with 67P properties. According to \cite{Jutzi2017} comets may have experienced several shape-changing collisions and the current shape of comet 67P would thus result from the last major shape-forming impact which most probably occurred within the last Gyr. Finally, \citet{schwartz2018catastrophic} suggests that catastrophic collisions may have occured and that the geological features of comets are not primordial.\\
Comets also interact with the interstellar medium (ISM) after their formation via two competing processes: accretion of ISM material and erosion caused by the impact of high-velocity ISM grains. According to simulations, the latter could be dominant for OC comets and could lead to a significant erosion of the cometary surface in the range of several meters \citep{Mumma1993}. However, erosion by ISM grains is a complex process depending for instance on the ISM grains composition and 3-dimensional structure and such erosion estimates are thus highly speculative \citep{Belyaev2010}.\\
One of the major candidates for altering cometary material are cosmic rays. Indeed, comets are constantly bombarded by Galactic Cosmic Rays (GCR). 
The bulk of the cosmic rays, including protons below 1~GeV, are able to penetrate into the surface layers of the comet \citep{Johnson1991}. 
However, higher energy cosmic rays penetrate deeper. In addition, comets are bombarded by stellar particles  at lower energies such as gamma ray bursts \citep{Johnson1991}.

The major goal of this paper is to quantify the energy deposited by cosmic rays and solar energetic particles in comets. In Section 2, we describe the CometCosmic model, based on Geant-4, that is at the core of this study. We also describe the modeled composition of the comet and the the Galactic Cosmic Rays (GCR) spectra used. In Section 3, the energy deposition in the comet by each source of particles is discussed both for KB and OC comets. We discuss the effect of the early solar system irradiation on the energy deposited in comets and the effect of the High-Z cosmic rays in Section 4. In a companion paper [Maggiolo et al. this issue, hereafter PART II], we will discuss the effects of energy deposition by GCR on the composition and structure of cometary nuclei, and how this can change our interpretation of space mission data.

 \begin{figure}
	 
	 
	 \noindent\includegraphics[width=45pc]{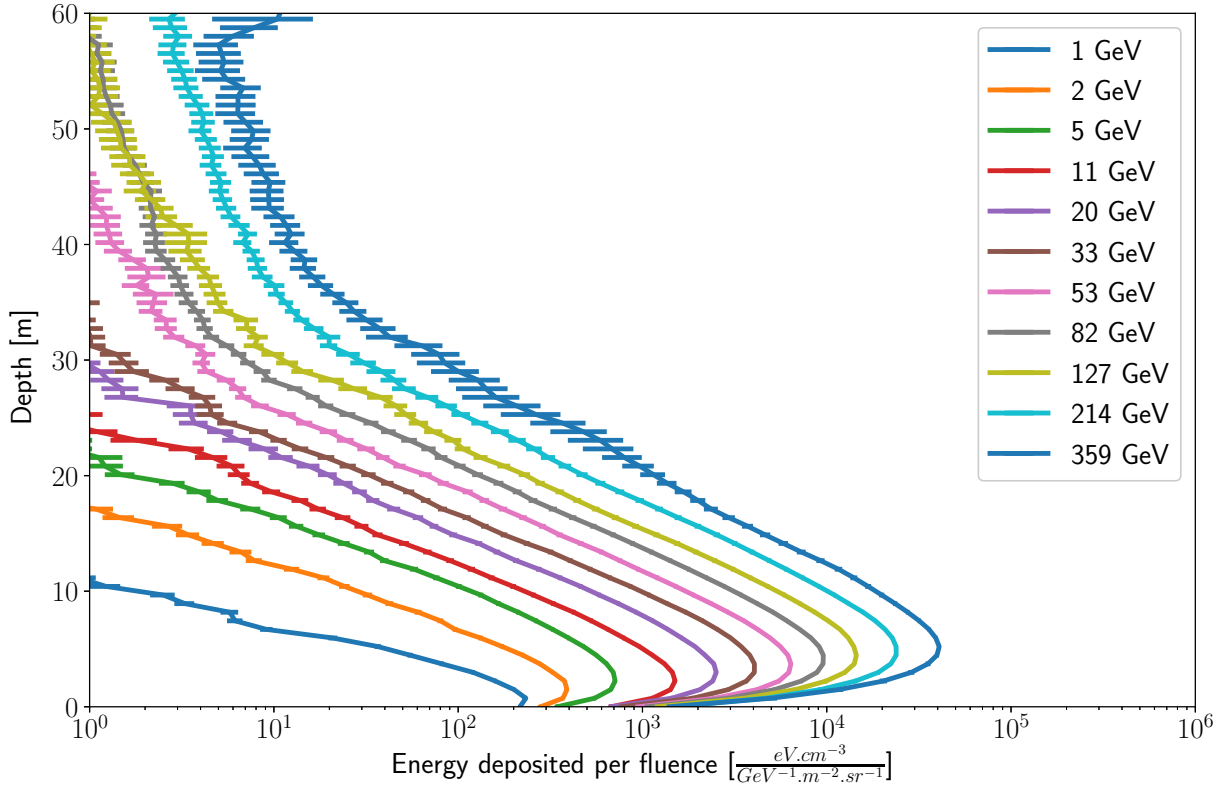}
 \caption{The deposition of energy by protons of different initial energy as a function of the depth and incident fluence in the comet nucleus. The more energetic the proton, the deeper the peak and the higher its intensity. Only a few initial proton energies are shown here, but the whole simulation considers a larger set of energies and more nuclei. The deposition is in dose per fluence, i.e. we have a dose in eV~cm$^{-3}$ if the fluence of a given particle, in GeV$^{-1}$m$^{-2}$sr$^{-1}$, is 1; this is equivalent to a dose per second per unit flux. }
 \label{figureGCR1}
 \end{figure}

 \begin{figure}
	
	 \noindent\includegraphics[width=40pc]{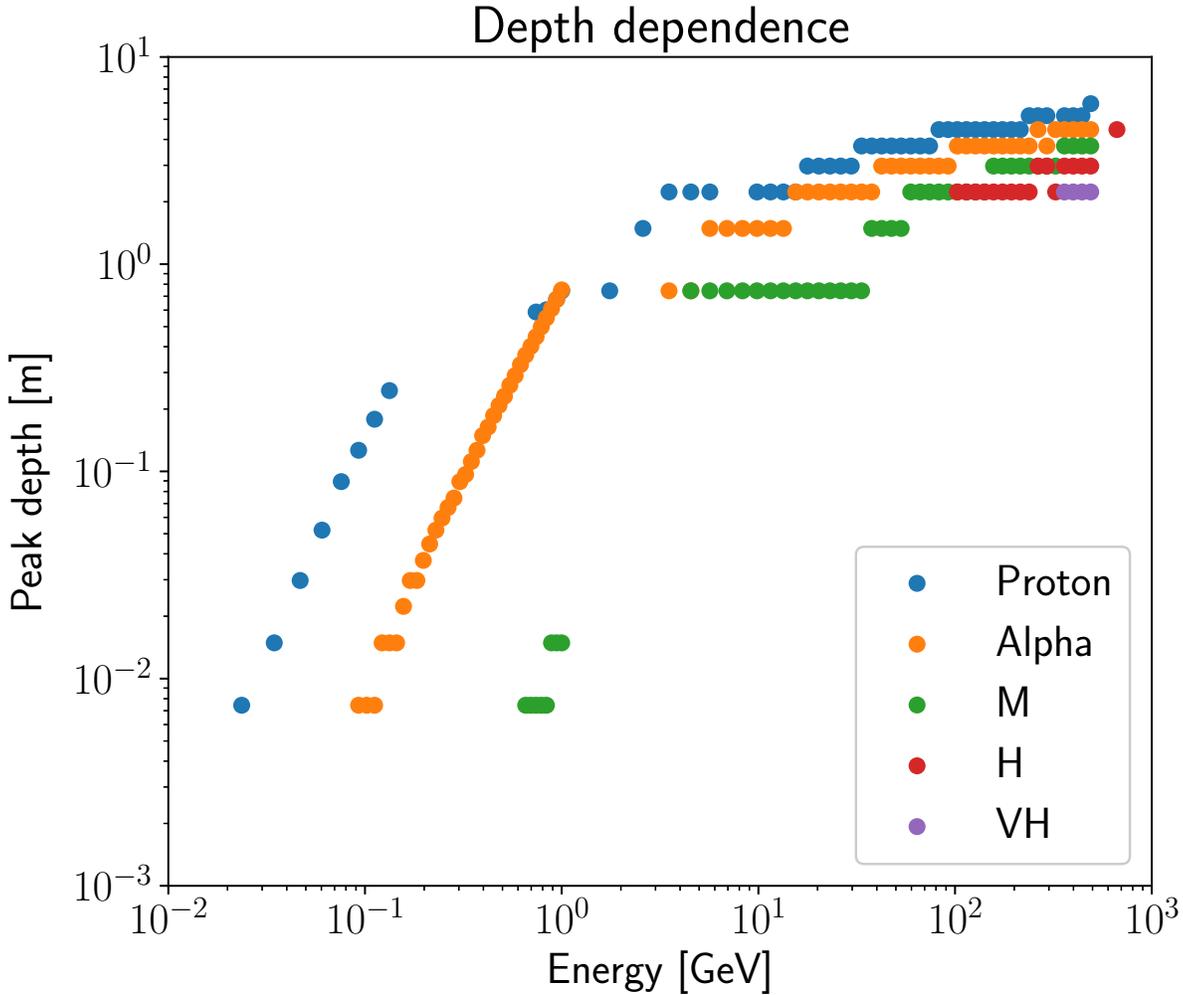}
 \caption{The peak energy deposition depth as a function of the energy of the primary particle computed by CometCosmic. Discontinuities are due to the resolution of the model, set at 1~cm depth below 1~GeV and 1~m above.}
 \label{figureGCR2}
 \end{figure}
\section{Model and Simulations}
\begin{figure}
	\noindent\includegraphics[width=45pc]{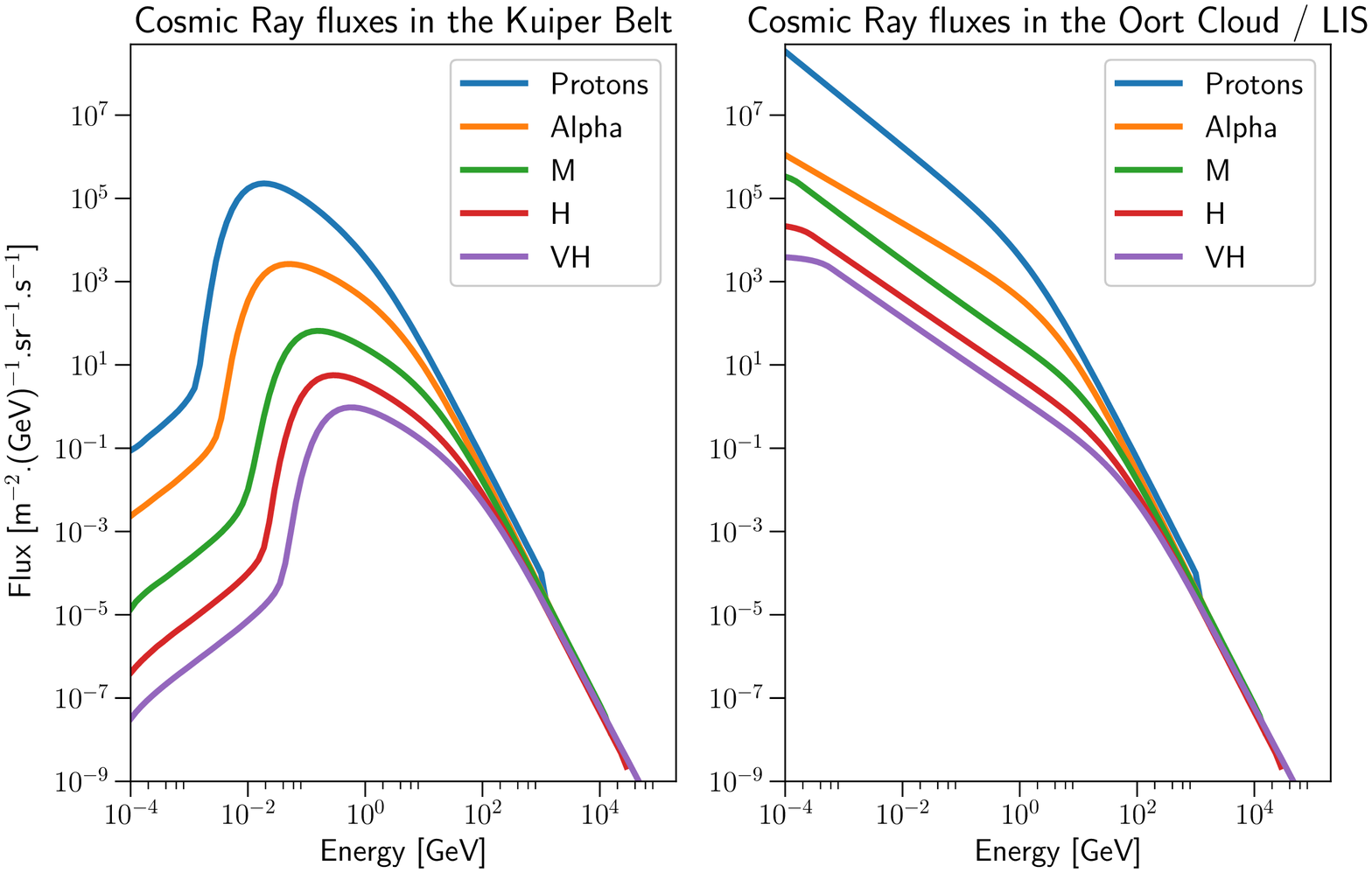}
 \caption{The Galactic Cosmic Ray flux in the Kuiper Belt and in the Oort Could (Local interstellar spectrum) as modeled by the Badhwar-O'Neill model. The solar modulation in the Kuiper Belt leads, as expected, to a large decrease of the flux at low energy. }
 \label{figureGCR3}
 \end{figure}
The simulation of the cosmic ray impact on comets was performed using CometCosmic, a new model based on the Geant-4 (GEometry ANd Tracking, \citet{AGOSTINELLI2003250}) library. For the present calculations, we used the version 4.9.6. This library allows to compute the transport of energetic particles --including the electromagnetic cascade, neutron creation and capture, and isotopic changes-- in a given medium.  Based on the recommendation for the library usage and on our previous experience with Geant-4 simulation \citep{gronoff_ionization_2009, sheel_numerical_2012,Gronoff2015,Gronoff2016}, we used the physics list QGSP\_BIC\_HP because it is the best suited to analyze the fragmentation of atomic nuclei and therefore to study the formation of isotopes. In addition, we modified the neutron model to be able to count the atomic nuclei created, allowing to fully evaluate the variation of the isotopic ratio due to cosmic rays \citep{Pavlov2014}.
We simulated a 5~km radius spherical comet, with density 0.5378 following the estimate for 67P \citep{Patzold2019}, composed of water ice (of isotopic composition $^{16}$O and $^1$H)  and SiO$_2$ with a dust (refractory material) /water ratio of 4 in mass (i.e. 5 H$_2$O for 6 SiO$_2$) \citep{Rotundiaaa3905}. This ratio has been further studied and is still debated; it has been estimated from higher than 6 \citep{stw1663} to lower than one \citep{sty3103} . For the study of the $^{15}$N variation (Paper II), we assume an initial N mixing ratio of 10$^{-3}$ \citep{Iro2003}; N is not included in the cosmic rays simulation to reduce the required amount of processor time. No complex surface material is assumed in this model, the refractory material (dust) is taken to be exclusively SiO$_2$, and the comet is considered as homogeneous, i.e. its porosity is homogeneous.(In actual comets, the heterogeneity leads to locations being more or less eroded, and leads to locations more pristine than other, which increases the chance of seeing GCR-weathering effects in observations).

For our simulations, the GCRs have been separated into different groups with a dedicated mass and charge to accelerate the computation \citep{Velinov2008}: Proton, Alpha, M group (Z=7, A=14 in our simulations, for a group that contains mainly carbon and oxygen nuclei), H group (Z=12, A =24, mainly silicon nuclei), and VH (Z = 26, A = 56,  mainly iron nuclei).  Please note that the energy is in GeV, and not in GeV/nucleon, to highlight the importance of the mass, and to be consistent with the other figures. We performed the Geant-4 simulations for each of these groups between 1~keV and 1~TeV, using two grids with exponentially distributed energies. The lower energies extend, i.e. 1~keV to 1~GeV allowed to show the influence of the solar wind --and interstellar particles-- on the first few centimeters of the comet, as well as the effects of the SEP. For that, 100 energies were selected on an exponentially increasing grid, and the products were computed on a 1~cm depth resolution grid. The higher energies, 1~GeV to 1~TeV, were optimized to the computation of the actual GCR and SEP influence; 50 energy bins were computed with a 1~m depth resolution comet.  The 1~TeV upper limit was determined empirically: above that energy, the increase in dose deposition is negligible in the comet, and while it can impact the lower depth (below 80~m) its overall effect on the comet lifetime (see Paper II) is negligible.  For all these energies and particle groups, 8 angles were taken to account for the spherical geometry of the comet. The number of particle shot in the Monte-Carlo simulation was optimized to have less than a 5\% uncertainty on the deposited dose at the peak, computed statistically. The error bars presented in the different figures have been computed using statistical analysis, and appropriately weighted \citep{gronoff_ionization_2009,gronoff:tel-00400638}.

Figure \ref{figureGCR1} shows the results of the simulation of energy deposition per unit volume in the comet due to  a proton flux (or fluence) irradiation of a given energy as a function of the depth inside the comet. This is an average profile since the depth depends upon the impact angle; our simulations involves integrating the isotropic flux over 8 angles for better depth accuracy, which is crucial in the spherical geometry \citep{gronoff_ionization_2011,norman_influence_2014, Gronoff2015}. The figure shows that energy deposition per unit volume peaks at a depth of 5~m for 300~GeV protons. 
Figure \ref{figureGCR2} shows the dependence of the peak production depth as a function of energy and precipitating particle type. This shows that, while particles with energy lower than 1~GeV have a peak energy deposition depth below 1~m, particles with higher energies penetrate deeper into the comet and deposit their energy in the first few tens of meter inside the nucleus. From 1 GeV to 100 GeV, while the incident energy varies by two orders of magnitude the peak energy deposition varies only a little (a few meters).

We used two cosmic ray flux models in our study, as shown in Figure \ref{figureGCR3}. The first one, the most realistic for 67P, considers the flux of cosmic rays in the KB region. The second one takes into account the Local Interstellar Spectra (LIS) to simulate the GCR flux in the OC region.

For computing these fluxes, we used the modified \citet{badhwar_improved_1992} model (hereafter B-O) H-BON10 \citep{mertens2013nairas}. While other modifications of B-O exist \citep[e.g.][]{matthia2013ready,o2014badhwar}, this is the only version of the B-O model that allows the radial distance to be specified as an input parameter. It also allows the solar modulation parameter to be parameterized by real-time, high-latitude neutron monitor data.  The solar modulation parameter for the H-BON10 model is shown in \citet{mertens2013nairas} over a 50-year period. For the calculations at 40 au, we used a solar modulation of 1000 MV so it reflects the average solar modulation. For the LIS, we used a 0 MV modulation; the distance having no impact in that case. The H-BON10 model gives the fluxes for all the ion; the different ions fluxes were added as a function of their charge to give the different group fluxes. M corresponds to ions from Z=3 to Z=10; H to from Z=11 to Z=20; and VH to Z $>$ 20. Our comparison for planetary atmospheres between planetocosmics and HZETRN \citep{gronoff_ionization_2011,Gronoff2015}, shows that this approximation is valid for our kind of study.

The difference in irradiation between a KB comet and an OC comet mainly results from the shielding by the heliospheric magnetic field. As a result, the flux of particles below $\sim$ 1 GeV is attenuated in the KB region.  The main assumption for this model is that the GCR flux has not significantly changed over 4.5~Gyr, and that the comet stayed in a region where the solar modulation of the GCR was stable \citep{2018A&A...618A..96P}. This very strong hypothesis will be discussed in Section \ref{earlysolar}. 
The anomalous cosmic ray flux \citep{giacalone_acceleration_2012,simnett_anomalous_2017} is also of interest for comet irradiation studies, their fluxes can be found in  \citet{Cummings2007}; however, when compared to the B-O model fluxes, they are of lower intensities and are therefore not included in our calculations.
The model does not take into account the irradiation by gamma ray bursts and solar photons as these particles/events have negligible effects below the first meter inside cometary nuclei according to our preliminary calculation using CometCosmic. 

\begin{figure}
	 \noindent\includegraphics[width=45pc]{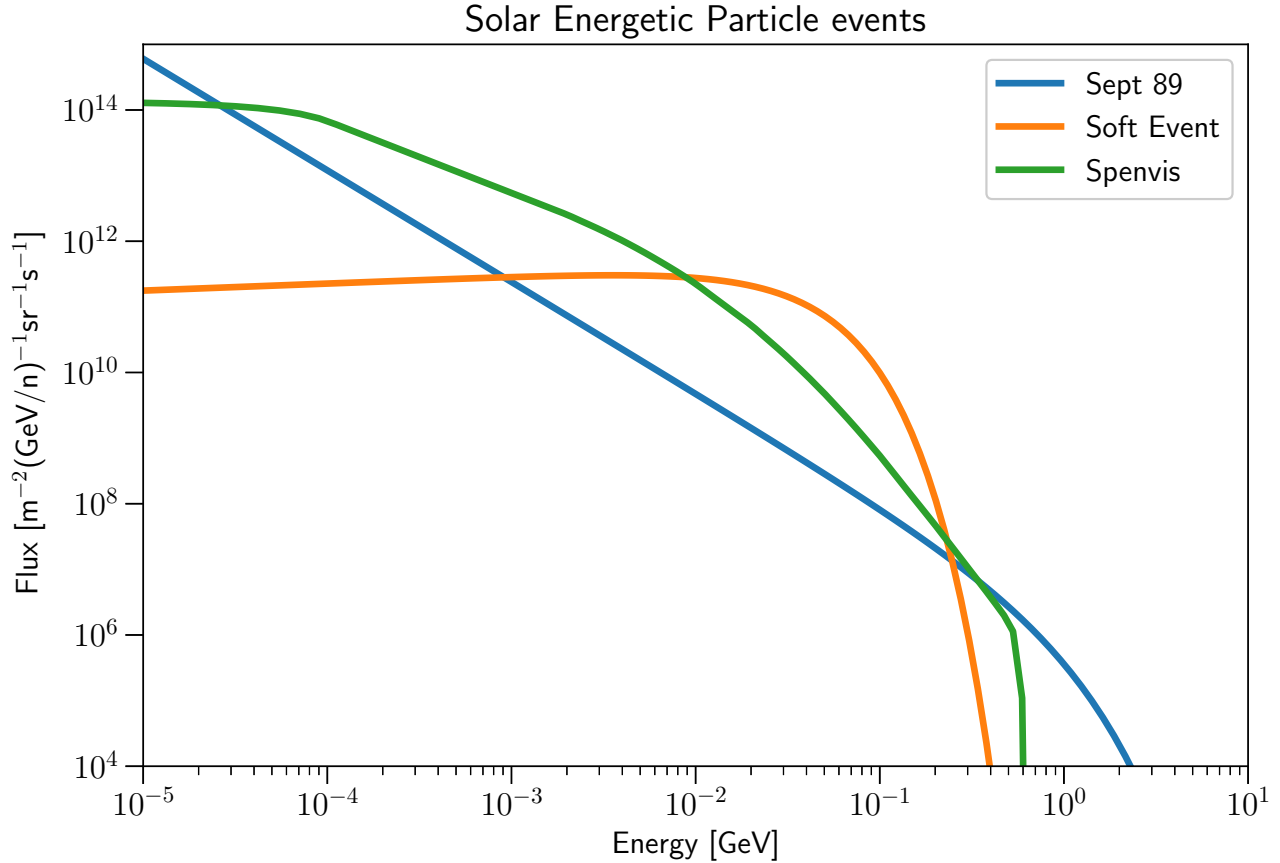}
 \caption{The Solar Energetic Particle events fluxes modelled in our calculations. These fluxes, which correspond to reconstructions of historical events, highlight different intensities and hardness (high energy tails) encountered in the different spectra. The Soft Event is based on the work of \citet{Smart2006} who tried to reconstruct a Carrington event. The validity of this spectra as a Carrington reconstruction has later been rejected, as discussed in \citet{Wolff2012,kovaltsov2014fluence}, and is used here as an example of an intense spectra with fewer particles above 100~MeV.  The ``SPENVIS'' event is the worst event of October 1989 as modeled by the European Space Agency's Space Environment Information System website \citet{SPENVIS2019}; the September 1989 event is based on the work of \citet{sheel_numerical_2012}.} 
 \label{figureSEP}
 \end{figure}

A second source of high-energy particles are the Solar Energetic Particle  (SEP) events. While they are rare, their influence over the  solar age has varied a lot \citep{Airapetian2016,fu2019}. In this paper, we study the impact that a SEP would have on a comet. Several SEPs flux models have been implemented in our model, taken from \citet{norman_influence_2014} and \citet{Gronoff2015} and are shown in Figure \ref{figureSEP}. These SEP event flux profiles were computed for a location at 1 au, which is far from the conditions encountered by the comet. If one considers a distance of 50 au, the reduction in flux is of the order of 2500, based on a R$^{-2}$ decrease (which is an approximation since magnetic fields can affect the dispersion with distance \citep{RodriguezGasen2014}). These SEP fluxes are comparable to those the comet must have suffered during the early phase of the solar system \citep{Airapetian2016}; therefore we use them for the present calculation. 

\section{Energy deposition in the comet}

There are many effects of dose deposition. For example, one can expect a 35~eV dose deposition to lead to ionization, provided that the particle depositing energy is at an energy above this threshold, which is the case in the present study. Ionization and dissociation/excitation of molecular species initiate chemical reactions and to the creation of molecular species not initially present in the comet. Several works on ice ionization [\citet{Johnson1991}, Paper II and references therein] have shown the chemical effects of dose deposition. One can expect a dose rate as small as 100~eV~cm$^{-3}$~s$^{-1}$ to create species in non-negligible quantities if sustained over billions of years as shown in Paper II.

\subsection{Galactic Cosmic Rays}

 \begin{figure}
	 \noindent\includegraphics{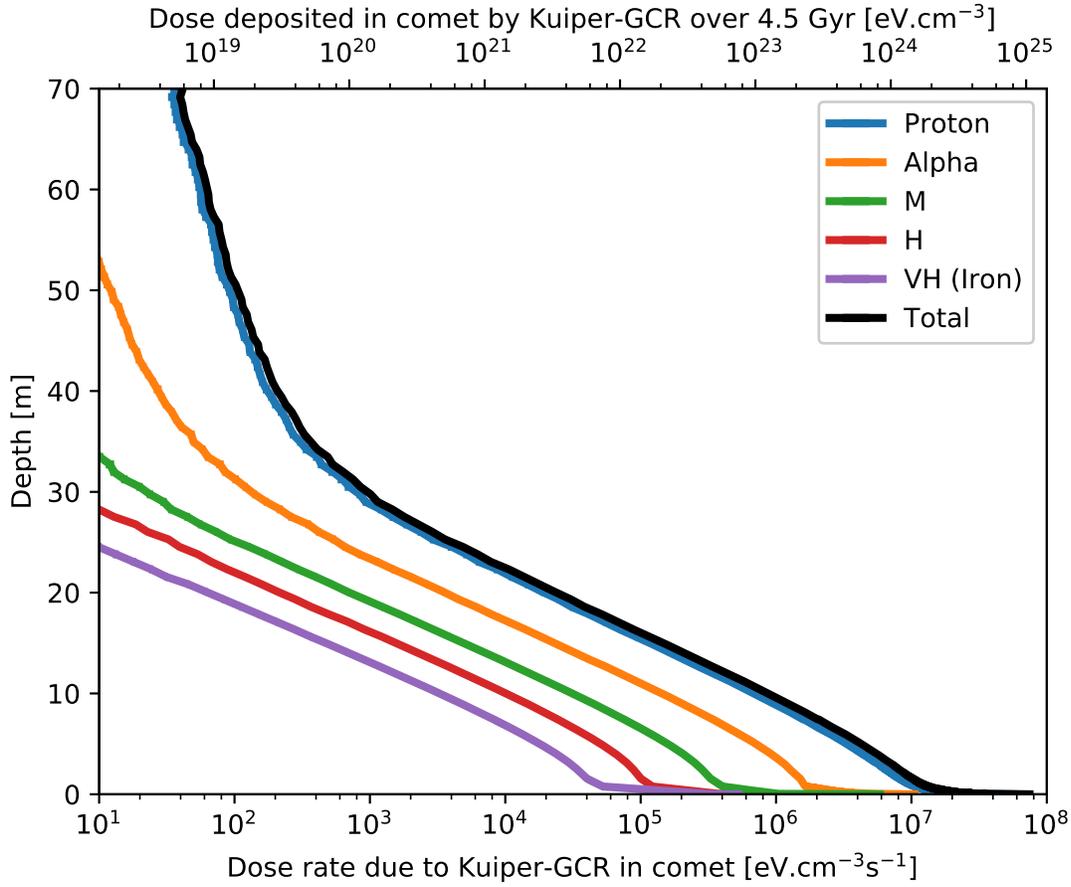}
 \caption{The dose deposited per second in the comet per cubic centimeter for the 67P conditions (in the Kuiper belt), in function of the incident GCR nuclei groups.}
 \label{figureDosePerSecond}
 \end{figure}
 \begin{figure}
	 \noindent\includegraphics{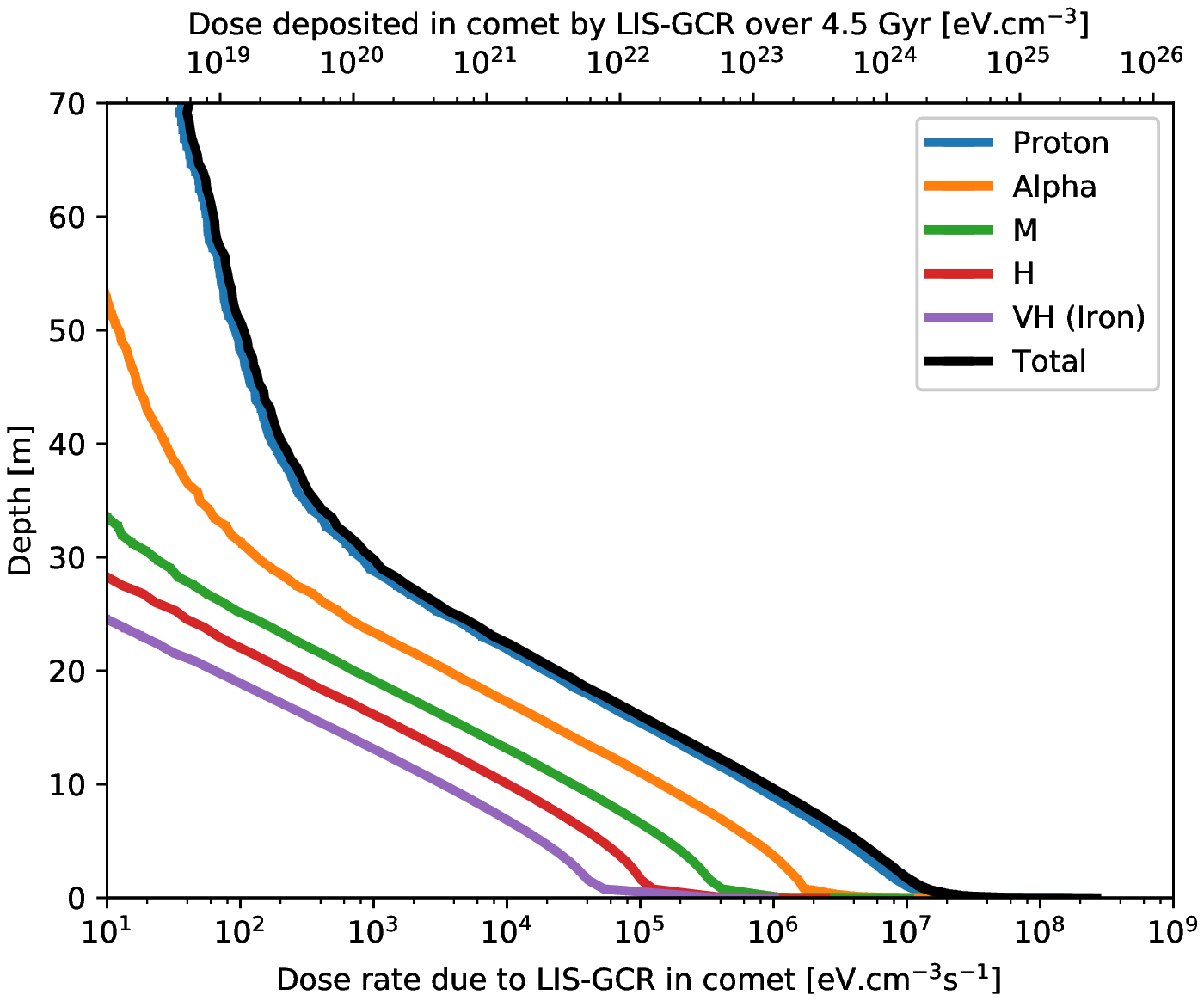}
	  
 \caption{The dose deposited per second in the comet per cubic centimeter for the Oort cloud (LIS) conditions, in function of the GCR nuclei group precipitating.}
 \label{figureDosePerSecondLIS}
 \end{figure}
 Figure \ref{figureDosePerSecond} presents the dose deposited in the comet per cubic centimeter per second for the conditions encountered in the KB.
 Figure \ref{figureDosePerSecondLIS} represents the dose deposited in the case of an OC comet. In both cases, the energetic protons are responsible for energy deposition up to 10$^8$ eV~cm$^{-3}$~s$^{-1}$ in the first centimeters. The difference between these two cases lies only in the first meter below the surface where for OC conditions more energy is deposited as, contrary to the KB region, the shielding of low energy GCR by the heliospheric magnetic field is negligible. This effect is demonstrated in Figure~\ref{figureDoseSEP}. In the following, we will therefore only show figures for the KB conditions. 
The effect of the refactory material/water ratio has an impact on the manner the energy will be dissipated in the comet: which kind of ionization and which kind of isotopes will be created. This ratio will therefore mainly change the effects modeled in paper II. The porosity of the comet will change the depth at which the energy deposition will happen: the depth could be changed into a column density as a first approximation to evaluate the effects of porosity, but the small size of the comet implies that spherical effects are quickly important if the porosity is changed in factors greater than $\approx$2.



 The cumulative impact after 4.5~Gyr can be seen at the top of the  Figures~\ref{figureDosePerSecond} and \ref{figureDosePerSecondLIS} (top x-axis). The dose computed is integrated over 4.5 Gyr by assuming that the cosmic ray flux in Figure~\ref{figureGCR1} does not vary in time. Therefore, a deposition dose rate of 10$^8$~eV~cm$^{-3}$~s$^{-1}$ corresponds to the energy deposition of $\approx$10$^{25}$~eV~cm$^{-3}$  This assumption will be discussed in more depth in Section \ref{earlysolar}.
 As expected, the protons produce most of the dose, with a deposition of the order of 10$^{25}$ eV~cm$^{-3}$ (i.e. 1.6~MJ~cm$^{-3}$) in the first meter below the surface. 


\subsection{Effects of High-Z}
\label{highz}

The alpha and high-Z cosmic rays (oxygen nuclei, carbon nuclei, and higher mass nuclei such as Fe) are a minority of the cosmic rays; however, their mass compensates their rarity \citep{gronoff_ionization_2011}, and the dissociation of nuclei can account for the creation of more isotopes. (Gronoff et al, in prep, adsr). As seen in Figures~\ref{figureDosePerSecond} and \ref{figureDosePerSecondLIS}, the dose deposition in the comet by the alpha and high-Z particles is small in comparison with proton deposition. 
These particles are however of interest for studying the deposition of heavy ions and the variation of isotopic ratio due to the different nuclear reactions (spallations, neutron capture) over the 4.5~Gyr lifetime of the comet.

\subsection{Additional Radiation Sources in the Early Solar System}
\label{earlysolar}
The question of the stability of the GCR spectra in time is a major unknown for studying space weather effects during eons. The amount of cosmic rays hitting the comet varied in time, notably because of the Early Sun's activity but also because of supernovae in the vicinity of the solar system. We used the modern spectra as an average for our study for the following reasons:

1) The early solar system is believed to have been exposed to more charged particle irradiation, as shows by meteoritic $^{26}$Al anomalies \citep{Feigelson1999,2018NatAs...2..709K}.
The major particle flux possible comes from the superflare associated coronal mass ejections that drive high fluence SEP events (also referred as solar cosmic rays in this context), with fluence in the order of 10$^{30}$ protons \citep{airapetian2019,fu2019}. Indeed, simulations shows that nearby supernovae could not explain the production of $^{26}$Al observed in the meteorites \citep{2018A&A...616A..85P}. The energies of such solar cosmic rays would be below 1~GeV \citep[][and references therein]{Airapetian2016,fu2019}, and therefore would be of negligible influence in the deeper layers of the comet. They affect only on the first meter below the surface. 

2) Considering explosion of different supernovae in the vicinity of the early solar system \citep{Torres2012}, it is difficult to estimate the corresponding total particle fluence. Most of the energy from such an event would be transported by particles below 1~MeV, and therefore would impact the first centimeter of the comet. Higher-energy particle would impact the comet like galactic cosmic rays but for a relatively small period of time. In any case, the first meter of a comet may be the first part of the comet to erode as it enters the inner solar system provided that it hasn't been eroded by high-velocity ISM grains earlier on. 
If the first layers were subject to particles from supernovae explosions, the change in isotopic ratio of non-diffusing species would mimic the GCR effects (because of similar spectra in the 1-100 GeV range), while if it were Sun related, the change would be skewed towards the first meter. The main advantage of doing such a study in a comet, with respect to an asteroid, is that a pristine comet would have been less affected by later solar events, and would have suffered fewer collisions. 


\subsection{Solar Wind and low-energy interstellar charged particles}

The solar wind is a continuous flow of ions and electrons, mostly containing H$^{+}$ ions (96 percent on average), alpha particles (4 percent on average) and a minor fraction of highly charged heavy ions \citep{2005ESASP.600E..44W}. Typical solar wind ions energy is of the order of 1 keV, and thus they deposit most of their energy in the first centimeter below the nucleus surface (see Figure \ref{figureGCR2}).\\
The average properties of the solar wind at heliocentric distance between 25 and 39 au has been estimated by \citet{Bagenal2015} using data from Voyager 2 (the measurements were made between 1988 and 1992). They obtain a median value of 0.0058 cm$^{-3}$ for the density and of 429 km.s$^{-1}$ for the velocity. Considering that these values are representative of solar wind conditions in the KB, we can estimate the solar wind flux reaching KB comets to be of the order of 2.4 10$^{7}$ eV~cm$^{-2}$~s$^{-1}$ (for the solar wind exclusively composed of protons). If we assume that all the solar wind energy is deposited in the first centimeter below the nucleus surface, the energy deposited per second in the nucleus of KB comets by the solar wind is approximately 2.4~10$^{7}$ eV~cm$^{-3}$~s$^{-1}$. Over the period of 4.5~Gy this corresponds to the total dose of 3.4~10$^{24}$~eV~cm$^{-3}$.\\
For OC comets, we use the solar wind properties from \citet{Bagenal2015} and consider that the solar wind propagates at constant velocity and that the density decreases with the square of the distance to the Sun. If the solar wind energy is deposited in the first centimeter below the surface it corresponds to a deposited energy of 68 eV~cm$^{-3}$~s$^{-1}$ (9.2 10$^{18}$ eV~cm$^{-3}$ over 4.5~Gy) at the inner edge of the OC (at 20000 au) and of 26 eV~cm$^{-3}$~s$^{-1}$ (3.7 10$^{17}$~eV~cm$^{-3}$ over 4.5~Gy) at the outer edge of the OC (at 100000 au).
Outside the heliosphere, the low-energy interstellar charged particle will also impact the comet. Such particles may interact with OC comets only as KB comets are located inside the heliosphere. Measurements by the Voyager probes show shown that the particle count rate drops outside of the heliopause except for energetic cosmic rays \citep[and references therein][]{krimigis2019energetic}. Considering the uncertainties on the spectrum of these particles, we do not compute the total energy deposited. However, the particle pressure in the heliosheath is dominated by protons with typical energies comprised between ~0.03 MeV and 4 MeV \citep[and references therein][]{krimigis2019energetic}. Owing to the low flux and low energy of interstellar plasma, it is possible to affirm that the deposition by the observed flux is negligible compared to the deposition by the GCR.

\subsection{Solar Energetic Particle events}

\begin{figure}
	 \noindent\includegraphics[width=20cm]{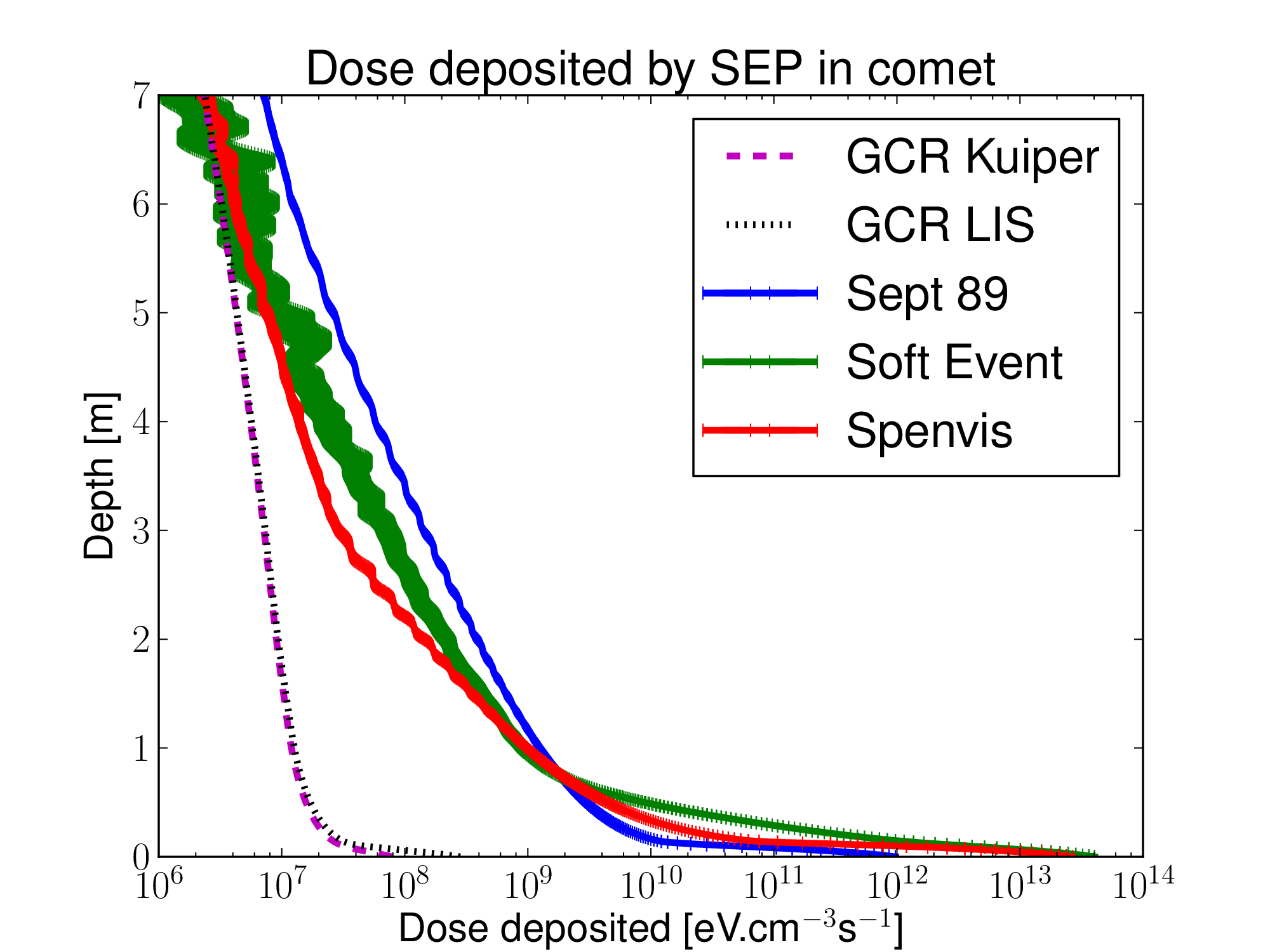}
 \caption{The dose deposited in the comet per cubic centimeter per second for the 67P conditions, in function of the SEP event shown in Figure~\ref{figureSEP}. (This corresponds to a Kuiper Belt comet subjected to a super-Carrington event or a comet in the inner Solar System subject to current SEP events). 
 The GCR doses are displayed to show how they become the dominant ionization source, in all conditions, deeper than $\approx$~8~m. 
 }
 \label{figureDoseSEP}
 \end{figure}

We simulated the energy deposited during an SEP event. We considered a super-Carrington SEP event at the location of the KB, which is what would be observed in the early solar system \citep{Airapetian2016,fu2019}.
As seen in Figure~\ref{figureDoseSEP} our simulation shows that the energy deposition via SEP events from the young Sun is mainly in the first meter. During such events, the deposition rate is higher than the GCR one in the first 8 meters. In the first centimeters below the surface the deposition rate of SEP events is several orders of magnitude higher than the one of GCR. At a depth of one meter the deposition rate is approximately 2 orders of magnitude higher for a  a super-Carrington SEP event, impacting the comet located in the KB, than for GCR. 
For present day SEP events, the dose rate should be reduced by a factor of 10$^4$; this is a dose comparable to the GCR one during the event in the first centimeters below the surface and approximately two orders of magnitude lower at a depth of 1 meter. \\
SEP events with significant amount of particles above 500 MeV are relatively rare now, of the order of 2 per year. They last for a couple days at most  \citep{2017SoPh..292...10K} but may have been more frequent in the early solar system \citep{Saxena_2019,airapetian2019}.  The total dose they deposit in cometary nuclei is therefore only  significant in the first centimeters below the surface. Deeper into the nucleus, the total energy deposited by SEP events over the lifetime of comets is well below the total energy deposited by GCR.


\section{Conclusion}
Most of the ionizing energy coming from the Sun --EUV-XUV, electrons, solar wind fluxes-- is deposited in the first meter of a comet \citep{Johnson1991,Johnson1997}. Its contribution to the energy deposition in the first centimeters below the surface is significant. In the KB region the energy deposited by the solar wind in the first centimeter below the surface can be of the same order of magnitude as the energy deposited by GCR. The irradiation fluxes via solar EUV-XUV, Gamma Ray Bursts, supernovae, and energetic particles emitted by the Sun at its early stage is much more difficult to constrain as we do not have a precise knowledge of their fluence and/or occurrence frequency. These events deposit most of their energy in the first meter below the cometary surface; the energy deposition at greater depth is dominated by GCR. The outer surface of the comet below which such particles deposit most of their energy interacts with the ISM during the lifetime of the comet and it is not clear if this interaction will result in the accretion of ISM material or in the erosion of the nucleus surface. In addition, the first meter of a cometary nucleus is  quickly removed upon a comet's first close approach to the Sun. We present here the calulations for the dose deposited deeper inside the cometary nucleus, below the first meter. The particles that have a potential to penetrate deeper in the comet are the SEPs and the GCR. The GCR  deposit energy in non-negligible amounts in the first 70 meters below the surface. This can change our view of the dynamically young comets, doing their first orbits in the inner solar system and for which the coma is formed  from those outer layers which have received a substantial dose of energy. In addition, the knowledge of the dose deposited over time may help understand the evolution of the KB objects observed by missions such as New Horizons. The effects of the energy deposition results in changes in  isotopic composition and  cometary chemistry, as described in Paper II.

\acknowledgments
The work of GG and WBM was supported by the Living Breathing Planet project: NASA Astrobiology Institute grant NNX15AE05G. The work of GG and VA was supported by the NASA Exobiology grant 80NSSC17K0463.
Work at BIRA-IASB regarding energetic particle impacts was supported by the Solar-Terrestrial Centre of Excellence, while ROSINA data analysis was supported via PRODEX/ROSINA PEA 90020 and an Additional Researchers Grant (Ministerial Decree of 2014-12-19), thanks to the Belgian Science Policy Office. AG thanks FNRS for a FRIA research grant. 
MR acknowledges the State of Bern and the Swiss National Science Foundation (SNSF, 200020\_182418).
HG was supported by the Swedish National Space Agency grant 108/18
CJM was supported by the Advanced Exploration Systems Division within the NASA Human Exploration and Operations Mission Directorate. GG wants to thank LFG for useful discussions. The authors wish to thank an anonymous referee for useful comments.






\bibliographystyle{plainnat}
\bibliography{cometcosmic1.bib}

\begin{thebibliography}{78}
\providecommand{\natexlab}[1]{#1}
\providecommand{\url}[1]{\texttt{#1}}
\expandafter\ifx\csname urlstyle\endcsname\relax
  \providecommand{\doi}[1]{doi: #1}\else
  \providecommand{\doi}{doi: \begingroup \urlstyle{rm}\Url}\fi

\bibitem[Agostinelli et~al.(2003)Agostinelli, Allison, Amako, Apostolakis,
  Araujo, Arce, Asai, Axen, Banerjee, Barrand, Behner, Bellagamba, Boudreau,
  Broglia, Brunengo, Burkhardt, Chauvie, Chuma, Chytracek, Cooperman, Cosmo,
  Degtyarenko, Dell'Acqua, Depaola, Dietrich, Enami, Feliciello, Ferguson,
  Fesefeldt, Folger, Foppiano, Forti, Garelli, Giani, Giannitrapani, Gibin,
  Cadenas, González, Abril, Greeniaus, Greiner, Grichine, Grossheim, Guatelli,
  Gumplinger, Hamatsu, Hashimoto, Hasui, Heikkinen, Howard, Ivanchenko,
  Johnson, Jones, Kallenbach, Kanaya, Kawabata, Kawabata, Kawaguti, Kelner,
  Kent, Kimura, Kodama, Kokoulin, Kossov, Kurashige, Lamanna, Lampén, Lara,
  Lefebure, Lei, Liendl, Lockman, Longo, Magni, Maire, Medernach, Minamimoto,
  de~Freitas, Morita, Murakami, Nagamatu, Nartallo, Nieminen, Nishimura,
  Ohtsubo, Okamura, O'Neale, Oohata, Paech, Perl, Pfeiffer, Pia, Ranjard,
  Rybin, Sadilov, Salvo, Santin, Sasaki, Savvas, Sawada, Scherer, Sei,
  Sirotenko, Smith, Starkov, Stoecker, Sulkimo, Takahata, Tanaka, Tcherniaev,
  Tehrani, Tropeano, Truscott, Uno, Urban, Urban, Verderi, Walkden, Wander,
  Weber, Wellisch, Wenaus, Williams, Wright, Yamada, Yoshida, and
  Zschiesche]{AGOSTINELLI2003250}
S.~Agostinelli, J.~Allison, K.~Amako, J.~Apostolakis, H.~Araujo, P.~Arce,
  M.~Asai, D.~Axen, S.~Banerjee, G.~Barrand, F.~Behner, L.~Bellagamba,
  J.~Boudreau, L.~Broglia, A.~Brunengo, H.~Burkhardt, S.~Chauvie, J.~Chuma,
  R.~Chytracek, G.~Cooperman, G.~Cosmo, P.~Degtyarenko, A.~Dell'Acqua,
  G.~Depaola, D.~Dietrich, R.~Enami, A.~Feliciello, C.~Ferguson, H.~Fesefeldt,
  G.~Folger, F.~Foppiano, A.~Forti, S.~Garelli, S.~Giani, R.~Giannitrapani,
  D.~Gibin, J.J.~Gómez Cadenas, I.~González, G.~Gracia Abril, G.~Greeniaus,
  W.~Greiner, V.~Grichine, A.~Grossheim, S.~Guatelli, P.~Gumplinger,
  R.~Hamatsu, K.~Hashimoto, H.~Hasui, A.~Heikkinen, A.~Howard, V.~Ivanchenko,
  A.~Johnson, F.W. Jones, J.~Kallenbach, N.~Kanaya, M.~Kawabata, Y.~Kawabata,
  M.~Kawaguti, S.~Kelner, P.~Kent, A.~Kimura, T.~Kodama, R.~Kokoulin,
  M.~Kossov, H.~Kurashige, E.~Lamanna, T.~Lampén, V.~Lara, V.~Lefebure,
  F.~Lei, M.~Liendl, W.~Lockman, F.~Longo, S.~Magni, M.~Maire, E.~Medernach,
  K.~Minamimoto, P.~Mora de~Freitas, Y.~Morita, K.~Murakami, M.~Nagamatu,
  R.~Nartallo, P.~Nieminen, T.~Nishimura, K.~Ohtsubo, M.~Okamura, S.~O'Neale,
  Y.~Oohata, K.~Paech, J.~Perl, A.~Pfeiffer, M.G. Pia, F.~Ranjard, A.~Rybin,
  S.~Sadilov, E.~Di Salvo, G.~Santin, T.~Sasaki, N.~Savvas, Y.~Sawada,
  S.~Scherer, S.~Sei, V.~Sirotenko, D.~Smith, N.~Starkov, H.~Stoecker,
  J.~Sulkimo, M.~Takahata, S.~Tanaka, E.~Tcherniaev, E.~Safai Tehrani,
  M.~Tropeano, P.~Truscott, H.~Uno, L.~Urban, P.~Urban, M.~Verderi, A.~Walkden,
  W.~Wander, H.~Weber, J.P. Wellisch, T.~Wenaus, D.C. Williams, D.~Wright,
  T.~Yamada, H.~Yoshida, and D.~Zschiesche.
\newblock Geant4—a simulation toolkit.
\newblock \emph{Nuclear Instruments and Methods in Physics Research Section A:
  Accelerators, Spectrometers, Detectors and Associated Equipment},
  506\penalty0 (3):\penalty0 250 -- 303, 2003.
\newblock ISSN 0168-9002.
\newblock \doi{https://doi.org/10.1016/S0168-9002(03)01368-8}.
\newblock URL
  \url{http://www.sciencedirect.com/science/article/pii/S0168900203013688}.

\bibitem[{Airapetian} et~al.(2016){Airapetian}, {Glocer}, {Gronoff},
  {H{\'e}brard}, and {Danchi}]{Airapetian2016}
V.~S. {Airapetian}, A.~{Glocer}, G.~{Gronoff}, E.~{H{\'e}brard}, and
  W.~{Danchi}.
\newblock {Prebiotic chemistry and atmospheric warming of early Earth by an
  active young Sun}.
\newblock \emph{Nature Geoscience}, 9:\penalty0 452--455, Jun 2016.
\newblock \doi{10.1038/ngeo2719}.

\bibitem[Airapetian et~al.(2019)Airapetian, Barnes, Cohen, Collinson, Danchi,
  Dong, Del~Genio, France, Garcia-Sage, Glocer, Gopalswamy, Grenfell, Gronoff,
  G\"udel, Herbst, Henning, Jackman, Jin, Johnstone, Kaltenegger, Kay,
  Kobayashi, Kuang, Li, Lynch, L"uftinger, Luhmann, Maehara, Mlynczak, Notsu,
  Ramirez, Rugheimer, Scheucher, Schlieder, Shibata, Sousa-Silva, Stamenkovi'c,
  Strangeway, Usmanov, Vergados, Verkhoglyadova, Vidotto, Voytek, Way, Zank,
  and Yamashiki]{airapetian2019}
V.~S. Airapetian, R.~Barnes, O.~Cohen, G.~A. Collinson, W.~C. Danchi, C.~F.
  Dong, A.~D. Del~Genio, K.~France, K.~Garcia-Sage, A.~Glocer, N.~Gopalswamy,
  J.~L. Grenfell, G.~Gronoff, M.~G\"udel, K.~Herbst, W.~G. Henning, C.~H.
  Jackman, M.~Jin, C.~P. Johnstone, L.~Kaltenegger, C.~D. Kay, K.~Kobayashi,
  W.~Kuang, G.~Li, B.~J. Lynch, T.~L"uftinger, TJ~G. Luhmann, H.~Maehara, M.~G.
  Mlynczak, Y.~Notsu, R.~M. Ramirez, S.~Rugheimer, M.~Scheucher, J.~E.
  Schlieder, K.~Shibata, C.~Sousa-Silva, V.~Stamenkovi'c, R.~J. Strangeway,
  A.~V. Usmanov, P.~Vergados, O.~P. Verkhoglyadova, A.~A. Vidotto, M.~Voytek,
  M.~J. Way, G.~P. Zank, and Y.~Yamashiki.
\newblock Impact of space weather on climate and habitability of
  terrestrial-type exoplanets.
\newblock \emph{International Journal of Astrobiology}, page 1–59, July 2019.
\newblock \doi{10.1017/S1473550419000132}.

\bibitem[Altwegg et~al.(2015)Altwegg, Balsiger, Bar-Nun, Berthelier, Bieler,
  Bochsler, Briois, Calmonte, Combi, De~Keyser, Eberhardt, Fiethe, Fuselier,
  Gasc, Gombosi, Hansen, H\"assig, J\"ackel, Kopp, Korth, LeRoy, Mall, Marty,
  Mousis, Neefs, Owen, R\`eme, Rubin, S\'emon, Tzou, Waite, and
  Wurz]{Altwegg2015}
K.~Altwegg, H.~Balsiger, A.~Bar-Nun, J.~J. Berthelier, A.~Bieler, P.~Bochsler,
  C.~Briois, U.~Calmonte, M.~Combi, J.~De~Keyser, P.~Eberhardt, B.~Fiethe,
  S.~Fuselier, S.~Gasc, T.~I. Gombosi, K.C. Hansen, M.~H\"assig, A.~J\"ackel,
  E.~Kopp, A.~Korth, L.~LeRoy, U.~Mall, B.~Marty, O.~Mousis, E.~Neefs, T.~Owen,
  H.~R\`eme, M.~Rubin, T.~S\'emon, C.-Y. Tzou, H.~Waite, and P.~Wurz.
\newblock 67p/churyumov-gerasimenko, a jupiter family comet with a high d/h
  ratio.
\newblock \emph{Science}, 347\penalty0 (6220), 2015.
\newblock \doi{10.1126/science.1261952}.
\newblock URL
  \url{http://www.sciencemag.org/content/347/6220/1261952.abstract}.

\bibitem[Altwegg et~al.(2017)Altwegg, Balsiger, Berthelier, Bieler, Calmonte,
  Keyser, Fiethe, Fuselier, Gasc, Gombosi, Owen, Roy, Rubin, Sémon, and
  Tzou]{Altwegg2017}
K.~Altwegg, H.~Balsiger, J.~J. Berthelier, A.~Bieler, U.~Calmonte, J.~De
  Keyser, B.~Fiethe, S.~A. Fuselier, S.~Gasc, T.~I. Gombosi, T.~Owen, L.~Le
  Roy, M.~Rubin, T.~Sémon, and C.-Y. Tzou.
\newblock D<sub>2</sub>o and hds in the coma of
  67p/churyumov\&\#x2013;gerasimenko.
\newblock \emph{Philosophical Transactions of the Royal Society A:
  Mathematical, Physical and Engineering Sciences}, 375\penalty0
  (2097):\penalty0 20160253, 2017.
\newblock \doi{10.1098/rsta.2016.0253}.
\newblock URL
  \url{https://royalsocietypublishing.org/doi/abs/10.1098/rsta.2016.0253}.

\bibitem[Badhwar and {O'Neill}(1992)]{badhwar_improved_1992}
G.~D. Badhwar and P.~M. {O'Neill}.
\newblock An improved model of galactic cosmic radiation for space exploration
  missions.
\newblock \emph{{International Journal of Radiation Applications and
  Instrumentation. Part D. Nuclear Tracks and Radiation Measurements}},
  20\penalty0 (3):\penalty0 403--410, July 1992.
\newblock ISSN 1359-0189.
\newblock \doi{10.1016/1359-0189(92)90024-P}.
\newblock URL
  \url{http://www.sciencedirect.com/science/article/B6X49-46MJ07Y-72/2/d78b722aa79f25a95a1e37c486935f8d}.

\bibitem[Bagenal et~al.(2015)Bagenal, Delamere, Elliott, Hill, Lisse, McComas,
  McNutt~Jr., Richardson, Smith, and Strobel]{Bagenal2015}
F.~Bagenal, P.~A. Delamere, H.~A. Elliott, M.~E. Hill, C.~M. Lisse, D.~J.
  McComas, R.~L. McNutt~Jr., J.~D. Richardson, C.~W. Smith, and D.~F. Strobel.
\newblock Solar wind at 33 au setting bounds on the pluto interaction for new
  horizons.
\newblock \emph{Journal of Geophysical Research: Planets}, 120\penalty0
  (9):\penalty0 1497--1511, 2015.
\newblock \doi{10.1002/2015JE004880}.
\newblock URL
  \url{https://agupubs.onlinelibrary.wiley.com/doi/abs/10.1002/2015JE004880}.

\bibitem[Balsiger et~al.(2015)Balsiger, Altwegg, Bar-Nun, Berthelier, Bieler,
  Bochsler, Briois, Calmonte, Combi, De~Keyser, et~al.]{Balsiger2015}
Hans Balsiger, Kathrin Altwegg, Akiva Bar-Nun, Jean-Jacques Berthelier, Andre
  Bieler, Peter Bochsler, Christelle Briois, Ursina Calmonte, Michael Combi,
  Johan De~Keyser, et~al.
\newblock {Detection of argon in the coma of comet 67P/Churyumov-Gerasimenko}.
\newblock \emph{Science advances}, 1\penalty0 (8):\penalty0 e1500377, 2015.

\bibitem[Bardyn et~al.(2017)Bardyn, Baklouti, Cottin, Fray, Briois, Paquette,
  Stenzel, Engrand, Fischer, Hornung, et~al.]{bardyn2017carbon}
Ana{\"\i}s Bardyn, Donia Baklouti, Herv{\'e} Cottin, Nicolas Fray, Christelle
  Briois, John Paquette, Oliver Stenzel, C{\'e}cile Engrand, Henning Fischer,
  Klaus Hornung, et~al.
\newblock Carbon-rich dust in comet 67p/churyumov-gerasimenko measured by
  cosima/rosetta.
\newblock \emph{Monthly Notices of the Royal Astronomical Society},
  469\penalty0 (Suppl\_2):\penalty0 S712--S722, 2017.

\bibitem[{Belyaev} and {Rafikov}(2010)]{Belyaev2010}
M.~A. {Belyaev} and R.~R. {Rafikov}.
\newblock {The Dynamics of Dust Grains in the Outer Solar System}.
\newblock \emph{\apj}, 723:\penalty0 1718--1735, November 2010.
\newblock \doi{10.1088/0004-637X/723/2/1718}.

\bibitem[Bertaux(2015)]{Bertaux2015}
Jean-Loup Bertaux.
\newblock Estimate of the erosion rate from h2o mass-loss measurements from
  swan/soho in previous perihelions of comet 67p/churyumov-gerasimenko and
  connection with observed rotation rate variations.
\newblock \emph{A\&A}, 583:\penalty0 A38, 2015.
\newblock \doi{10.1051/0004-6361/201525992}.
\newblock URL \url{https://doi.org/10.1051/0004-6361/201525992}.

\bibitem[{Bieler} et~al.(2015){Bieler}, {Altwegg}, and {Balsiger}]{Bieler2015}
A.~{Bieler}, K.~{Altwegg}, and et~al. {Balsiger}.
\newblock {Abundant molecular oxygen in the coma of comet
  67P/Churyumov-Gerasimenko}.
\newblock \emph{Nature}, 526:\penalty0 678--681, October 2015.
\newblock \doi{10.1038/nature15707}.

\bibitem[Capaccioni et~al.(2015)Capaccioni, Coradini, Filacchione, Erard,
  Arnold, Drossart, De~Sanctis, Bockelee-Morvan, Capria, Tosi, Leyrat, Schmitt,
  Quirico, Cerroni, Mennella, Raponi, Ciarniello, McCord, Moroz, Palomba,
  Ammannito, Barucci, Bellucci, Benkhoff, Bibring, Blanco, Blecka, Carlson,
  Carsenty, Colangeli, Combes, Combi, Crovisier, Encrenaz, Federico, Fink,
  Fonti, Ip, Irwin, Jaumann, Kuehrt, Langevin, Magni, Mottola, Orofino,
  Palumbo, Piccioni, Schade, Taylor, Tiphene, Tozzi, Beck, Biver, Bonal, Combe,
  Despan, Flamini, Fornasier, Frigeri, Grassi, Gudipati, Longobardo, Markus,
  Merlin, Orosei, Rinaldi, Stephan, Cartacci, Cicchetti, Giuppi, Hello, Henry,
  Jacquinod, Noschese, Peter, Politi, Reess, and Semery]{Capaccioniaaa0628}
F.~Capaccioni, A.~Coradini, G.~Filacchione, S.~Erard, G.~Arnold, P.~Drossart,
  M.~C. De~Sanctis, D.~Bockelee-Morvan, M.~T. Capria, F.~Tosi, C.~Leyrat,
  B.~Schmitt, E.~Quirico, P.~Cerroni, V.~Mennella, A.~Raponi, M.~Ciarniello,
  T.~McCord, L.~Moroz, E.~Palomba, E.~Ammannito, M.~A. Barucci, G.~Bellucci,
  J.~Benkhoff, J.~P. Bibring, A.~Blanco, M.~Blecka, R.~Carlson, U.~Carsenty,
  L.~Colangeli, M.~Combes, M.~Combi, J.~Crovisier, T.~Encrenaz, C.~Federico,
  U.~Fink, S.~Fonti, W.~H. Ip, P.~Irwin, R.~Jaumann, E.~Kuehrt, Y.~Langevin,
  G.~Magni, S.~Mottola, V.~Orofino, P.~Palumbo, G.~Piccioni, U.~Schade,
  F.~Taylor, D.~Tiphene, G.~P. Tozzi, P.~Beck, N.~Biver, L.~Bonal, J.-Ph.
  Combe, D.~Despan, E.~Flamini, S.~Fornasier, A.~Frigeri, D.~Grassi,
  M.~Gudipati, A.~Longobardo, K.~Markus, F.~Merlin, R.~Orosei, G.~Rinaldi,
  K.~Stephan, M.~Cartacci, A.~Cicchetti, S.~Giuppi, Y.~Hello, F.~Henry,
  S.~Jacquinod, R.~Noschese, G.~Peter, R.~Politi, J.~M. Reess, and A.~Semery.
\newblock The organic-rich surface of comet 67p/churyumov-gerasimenko as seen
  by virtis/rosetta.
\newblock \emph{Science}, 347\penalty0 (6220), 2015.
\newblock ISSN 0036-8075.
\newblock \doi{10.1126/science.aaa0628}.
\newblock URL \url{https://science.sciencemag.org/content/347/6220/aaa0628}.

\bibitem[{Carusi} et~al.(1985){Carusi}, {Kresak}, {Perozzi}, and
  {Valsecchi}]{1985ltes.book.....C}
A.~{Carusi}, L.~{Kresak}, E.~{Perozzi}, and G.~B. {Valsecchi}.
\newblock \emph{{Long-term evolution of short-period comets}}.
\newblock Bristol, England and Accord, MA, Adam Hilger, Ltd., 1985, 272 p.,
  1985.

\bibitem[{Cummings} and {Stone}(2007)]{Cummings2007}
A.~C. {Cummings} and E.~C. {Stone}.
\newblock {Composition of Anomalous Cosmic Rays}.
\newblock \emph{\ssr}, 130:\penalty0 389--399, June 2007.
\newblock \doi{10.1007/s11214-007-9161-y}.

\bibitem[{Davidsson} et~al.(2016){Davidsson}, {Sierks}, {G{\"u}ttler},
  {Marzari}, {Pajola}, {Rickman}, {A'Hearn}, {Auger}, {El-Maarry}, {Fornasier},
  {Guti{\'e}rrez}, {Keller}, {Massironi}, {Snodgrass}, {Vincent}, {Barbieri},
  {Lamy}, {Rodrigo}, {Koschny}, {Barucci}, {Bertaux}, {Bertini}, {Cremonese},
  {Da Deppo}, {Debei}, {De Cecco}, {Feller}, {Fulle}, {Groussin}, {Hviid},
  {H{\"o}fner}, {Ip}, {Jorda}, {Knollenberg}, {Kovacs}, {Kramm}, {K{\"u}hrt},
  {K{\"u}ppers}, {La Forgia}, {Lara}, {Lazzarin}, {Lopez Moreno},
  {Moissl-Fraund}, {Mottola}, {Naletto}, {Oklay}, {Thomas}, and
  {Tubiana}]{Davidsson2016}
B.~J.~R. {Davidsson}, H.~{Sierks}, C.~{G{\"u}ttler}, F.~{Marzari}, M.~{Pajola},
  H.~{Rickman}, M.~F. {A'Hearn}, A.-T. {Auger}, M.~R. {El-Maarry},
  S.~{Fornasier}, P.~J. {Guti{\'e}rrez}, H.~U. {Keller}, M.~{Massironi},
  C.~{Snodgrass}, J.-B. {Vincent}, C.~{Barbieri}, P.~L. {Lamy}, R.~{Rodrigo},
  D.~{Koschny}, M.~A. {Barucci}, J.-L. {Bertaux}, I.~{Bertini}, G.~{Cremonese},
  V.~{Da Deppo}, S.~{Debei}, M.~{De Cecco}, C.~{Feller}, M.~{Fulle},
  O.~{Groussin}, S.~F. {Hviid}, S.~{H{\"o}fner}, W.-H. {Ip}, L.~{Jorda},
  J.~{Knollenberg}, G.~{Kovacs}, J.-R. {Kramm}, E.~{K{\"u}hrt},
  M.~{K{\"u}ppers}, F.~{La Forgia}, L.~M. {Lara}, M.~{Lazzarin}, J.~J. {Lopez
  Moreno}, R.~{Moissl-Fraund}, S.~{Mottola}, G.~{Naletto}, N.~{Oklay},
  N.~{Thomas}, and C.~{Tubiana}.
\newblock {The primordial nucleus of comet 67P/Churyumov-Gerasimenko}.
\newblock \emph{\aap}, 592:\penalty0 A63, July 2016.
\newblock \doi{10.1051/0004-6361/201526968}.

\bibitem[{De Keyser} et~al.(2017){De Keyser}, Dhooghe, Altwegg, Balsiger,
  Berthelier, Briois, Calmonte, Cessateur, Combi, Equeter, Fiethe, Fuselier,
  Gasc, Gibbons, Gombosi, Gunell, H\"{a}ssig, {Le Roy}, Maggiolo, Mall, Marty,
  Neefs, R\`{e}me, Rubin, S\'{e}mon, Tzou, and Wurz]{DeKeyserEtAl2017}
J.~{De Keyser}, F.~Dhooghe, K.~Altwegg, H.~Balsiger, J.-J. Berthelier, Ch.
  Briois, U.~Calmonte, G.~Cessateur, M.R. Combi, E.~Equeter, B.~Fiethe,
  S.~Fuselier, S.~Gasc, A.~Gibbons, T.~Gombosi, H.~Gunell, M.~H\"{a}ssig,
  L.~{Le Roy}, R.~Maggiolo, U.~Mall, B.~Marty, E.~Neefs, H.~R\`{e}me, M.~Rubin,
  T.~S\'{e}mon, C.-Y. Tzou, and P.~Wurz.
\newblock Evidence for distributed gas sources of hydrogen halides in the coma
  of comet 67p/churyumov-gerasimenko.
\newblock \emph{MNRAS}, 469:\penalty0 S695–S711, 2017.
\newblock \doi{10.1093/mnras/stx2725}.

\bibitem[Dhooghe et~al.(2017)Dhooghe, {De Keyser}, Altwegg, Briois, Balsiger,
  Berthelier, Calmonte, Cessateur, Combi, Equeter, Fiethe, Fray, Fuselier,
  Gasc, Gibbons, Gombosi, Gunell, H\"{a}ssig, Hilchenbach, {Le Roy}, Maggiolo,
  Mall, Marty, Neefs, R\`{e}me, Rubin, S\'{e}mon, Tzou, and
  Wurz]{DhoogheEtAl2017}
F.~Dhooghe, J.~{De Keyser}, K.~Altwegg, Ch. Briois, H.~Balsiger, J.-J.
  Berthelier, U.~Calmonte, G.~Cessateur, M.~R. Combi, E.~Equeter, B.~Fiethe,
  N.~Fray, S.~Fuselier, S.~Gasc, A.~Gibbons, T.~Gombosi, H.~Gunell,
  M.~H\"{a}ssig, M.~Hilchenbach, L.~{Le Roy}, R.~Maggiolo, U.~Mall, B.~Marty,
  E.~Neefs, H.~R\`{e}me, M.~Rubin, T.~S\'{e}mon, C.-Y. Tzou, and P.~Wurz.
\newblock {Halogens as tracers of protosolar nebula material in comet
  67P/Churyumov-Gerasimenko}.
\newblock \emph{MNRAS}, 472:\penalty0 1336–1345, 2017.
\newblock \doi{10.1093/mnras/stx1911}.

\bibitem[{Duncan} et~al.(2004){Duncan}, {Levison}, and {Dones}]{Duncan2004}
M.~{Duncan}, H.~{Levison}, and L.~{Dones}.
\newblock \emph{{Dynamical evolution of ecliptic comets}}, pages 193--204.
\newblock 2004.

\bibitem[Feigelson and Montmerle(1999)]{Feigelson1999}
Eric~D. Feigelson and Thierry Montmerle.
\newblock High-energy processes in young stellar objects.
\newblock \emph{Annual Review of Astronomy and Astrophysics}, 37\penalty0
  (1):\penalty0 363--408, 1999.
\newblock \doi{10.1146/annurev.astro.37.1.363}.
\newblock URL \url{https://doi.org/10.1146/annurev.astro.37.1.363}.

\bibitem[Fu et~al.(2019)Fu, Jiang, Airapetian, Hu, Li, and Zank]{fu2019}
Shuai Fu, Yong Jiang, Vladimir Airapetian, Junxiang Hu, Gang Li, and Gary Zank.
\newblock Effect of star rotation rate on the characteristics of energetic
  particle events.
\newblock \emph{The Astrophysical Journal Letters}, 878\penalty0 (2):\penalty0
  L36, 2019.

\bibitem[Fulle et~al.(2016)Fulle, Altobelli, Buratti, Choukroun, Fulchignoni,
  Grün, Taylor, and Weissman]{stw1663}
Marco Fulle, N.~Altobelli, B.~Buratti, M.~Choukroun, M.~Fulchignoni, E.~Grün,
  M.~G. G.~T. Taylor, and P.~Weissman.
\newblock {Unexpected and significant findings in comet
  67P/Churyumov–Gerasimenko: an interdisciplinary view}.
\newblock \emph{Monthly Notices of the Royal Astronomical Society},
  462\penalty0 (Suppl 1):\penalty0 S2--S8, 09 2016.
\newblock ISSN 0035-8711.
\newblock \doi{10.1093/mnras/stw1663}.
\newblock URL \url{https://doi.org/10.1093/mnras/stw1663}.

\bibitem[Giacalone et~al.(2012)Giacalone, Drake, and
  Jokipii]{giacalone_acceleration_2012}
J.~Giacalone, J.~F. Drake, and J.~R. Jokipii.
\newblock The {Acceleration} {Mechanism} of {Anomalous} {Cosmic} {Rays}.
\newblock \emph{Space Science Reviews}, 173\penalty0 (1-4):\penalty0 283--307,
  November 2012.
\newblock ISSN 0038-6308, 1572-9672.
\newblock \doi{{10.1007/s11214-012-9915-z}}.
\newblock URL \url{http://link.springer.com/article/10.1007/s11214-012-9915-z}.

\bibitem[Gronoff et~al.(2009)Gronoff, Lilensten, Desorgher, and
  Fl\"uckiger]{gronoff_ionization_2009}
G.~Gronoff, J.~Lilensten, L.~Desorgher, and E.~Fl\"uckiger.
\newblock Ionization processes in the atmosphere of titan. i. ionization in the
  whole atmosphere.
\newblock \emph{\aap}, 506:\penalty0 955--964, November 2009.

\bibitem[Gronoff et~al.(2011)Gronoff, Mertens, Lilensten, Desorgher,
  Flückiger, and Velinov]{gronoff_ionization_2011}
G.~Gronoff, C.~Mertens, J.~Lilensten, L.~Desorgher, E.~Flückiger, and
  P.~Velinov.
\newblock Ionization processes in the atmosphere of titan. {III.} ionization by
  high-z nuclei cosmic rays.
\newblock \emph{\aap}, 529:\penalty0 143, May 2011.

\bibitem[{Gronoff} et~al.(2015){Gronoff}, {Norman}, and {Mertens}]{Gronoff2015}
G.~{Gronoff}, R.~B. {Norman}, and C.~J. {Mertens}.
\newblock {Computation of cosmic ray ionization and dose at Mars. I: A
  comparison of HZETRN and Planetocosmics for proton and alpha particles}.
\newblock \emph{Advances in Space Research}, 55:\penalty0 1799--1805, April
  2015.
\newblock \doi{10.1016/j.asr.2015.01.028}.

\bibitem[{Gronoff} et~al.(2016){Gronoff}, {Mertens}, {Norman}, {Straume}, and
  {Lusby}]{Gronoff2016}
G.~{Gronoff}, C.~J. {Mertens}, R.~B. {Norman}, T.~{Straume}, and T.~C. {Lusby}.
\newblock {Assessment of the influence of the RaD-X balloon payload on the
  onboard radiation detectors}.
\newblock \emph{Space Weather}, 14:\penalty0 835--845, October 2016.
\newblock \doi{10.1002/2016SW001405}.

\bibitem[Gronoff(2009)]{gronoff:tel-00400638}
Guillaume Gronoff.
\newblock \emph{{Energetic inputs in planetary atmospheres. Titan, Venus and
  Mars.}}
\newblock Theses, {Universit{\'e} Joseph-Fourier - Grenoble I}, June 2009.
\newblock URL \url{https://tel.archives-ouvertes.fr/tel-00400638}.

\bibitem[Groussin et~al.(2015)Groussin, {Sierks, H.}, {Barbieri, C.}, {Lamy,
  P.}, {Rodrigo, R.}, {Koschny, D.}, {Rickman, H.}, {Keller, H. U.},
  {A\'{}Hearn, M. F.}, {Auger, A.-T.}, {Barucci, M. A.}, {Bertaux, J.-L.},
  {Bertini, I.}, {Besse, S.}, {Cremonese, G.}, {Da Deppo, V.}, {Davidsson, B.},
  {Debei, S.}, {De Cecco, M.}, {El-Maarry, M. R.}, {Fornasier, S.}, {Fulle,
  M.}, {Guti\'errez, P. J.}, {G\"uttler, C.}, {Hviid, S.}, {Ip, W.-H}, {Jorda,
  L.}, {Knollenberg, J.}, {Kovacs, G.}, {Kramm, J. R.}, {K\"uhrt, E.},
  {K\"uppers, M.}, {Lara, L. M.}, {Lazzarin, M.}, {Lopez Moreno, J. J.},
  {Lowry, S.}, {Marchi, S.}, {Marzari, F.}, {Massironi, M.}, {Mottola, S.},
  {Naletto, G.}, {Oklay, N.}, {Pajola, M.}, {Pommerol, A.}, {Thomas, N.},
  {Toth, I.}, {Tubiana, C.}, and {Vincent, J.-B.}]{Groussin2015}
O.~Groussin, {Sierks, H.}, {Barbieri, C.}, {Lamy, P.}, {Rodrigo, R.}, {Koschny,
  D.}, {Rickman, H.}, {Keller, H. U.}, {A\'{}Hearn, M. F.}, {Auger, A.-T.},
  {Barucci, M. A.}, {Bertaux, J.-L.}, {Bertini, I.}, {Besse, S.}, {Cremonese,
  G.}, {Da Deppo, V.}, {Davidsson, B.}, {Debei, S.}, {De Cecco, M.},
  {El-Maarry, M. R.}, {Fornasier, S.}, {Fulle, M.}, {Guti\'errez, P. J.},
  {G\"uttler, C.}, {Hviid, S.}, {Ip, W.-H}, {Jorda, L.}, {Knollenberg, J.},
  {Kovacs, G.}, {Kramm, J. R.}, {K\"uhrt, E.}, {K\"uppers, M.}, {Lara, L. M.},
  {Lazzarin, M.}, {Lopez Moreno, J. J.}, {Lowry, S.}, {Marchi, S.}, {Marzari,
  F.}, {Massironi, M.}, {Mottola, S.}, {Naletto, G.}, {Oklay, N.}, {Pajola,
  M.}, {Pommerol, A.}, {Thomas, N.}, {Toth, I.}, {Tubiana, C.}, and {Vincent,
  J.-B.}
\newblock Temporal morphological changes in the imhotep region of comet
  67p/churyumov-gerasimenko.
\newblock \emph{A\&A}, 583:\penalty0 A36, 2015.
\newblock \doi{10.1051/0004-6361/201527020}.
\newblock URL \url{https://doi.org/10.1051/0004-6361/201527020}.

\bibitem[{Guilbert-Lepoutre} et~al.(2015){Guilbert-Lepoutre}, {Besse},
  {Mousis}, {Ali-Dib}, {H{\"o}fner}, {Koschny}, and
  {Hager}]{Guilbert-Lepoutre2015}
A.~{Guilbert-Lepoutre}, S.~{Besse}, O.~{Mousis}, M.~{Ali-Dib}, S.~{H{\"o}fner},
  D.~{Koschny}, and P.~{Hager}.
\newblock {On the Evolution of Comets}.
\newblock \emph{\ssr}, 197:\penalty0 271--296, December 2015.
\newblock \doi{10.1007/s11214-015-0148-9}.

\bibitem[{Iro} et~al.(2003){Iro}, {Gautier}, {Hersant}, {Bockel{\'e}e-Morvan},
  and {Lunine}]{Iro2003}
N.~{Iro}, D.~{Gautier}, F.~{Hersant}, D.~{Bockel{\'e}e-Morvan}, and J.~I.
  {Lunine}.
\newblock An interpretation of the nitrogen deficiency in comets.
\newblock \emph{Icarus}, 161:\penalty0 511--532, February 2003.
\newblock \doi{10.1016/S0019-1035(02)00038-6}.

\bibitem[{Johnson}(1991)]{Johnson1991}
R.~E. {Johnson}.
\newblock Irradiation effects in a comet's outer layers.
\newblock \emph{\jgr}, 96:\penalty0 17553, September 1991.
\newblock \doi{10.1029/91JE01743}.

\bibitem[{Johnson} and {Quickenden}(1997)]{Johnson1997}
R.~E. {Johnson} and T.~I. {Quickenden}.
\newblock Photolysis and radiolysis of water ice on outer solar system bodies.
\newblock \emph{\jgr}, 102:\penalty0 10985--10996, 1997.
\newblock \doi{10.1029/97JE00068}.

\bibitem[{Jutzi} and {Benz}(2017)]{JutziBenz2017}
M.~{Jutzi} and W.~{Benz}.
\newblock {Formation of bi-lobed shapes by sub-catastrophic collisions. A late
  origin of comet 67P's structure}.
\newblock \emph{\aap}, 597:\penalty0 A62, January 2017.
\newblock \doi{10.1051/0004-6361/201628964}.

\bibitem[{Jutzi} et~al.(2017){Jutzi}, {Benz}, {Toliou}, {Morbidelli}, and
  {Brasser}]{Jutzi2017}
M.~{Jutzi}, W.~{Benz}, A.~{Toliou}, A.~{Morbidelli}, and R.~{Brasser}.
\newblock {How primordial is the structure of comet 67P?. Combined collisional
  and dynamical models suggest a late formation}.
\newblock \emph{\aap}, 597:\penalty0 A61, January 2017.
\newblock \doi{10.1051/0004-6361/201628963}.

\bibitem[Keller et~al.(2015)Keller, {Mottola, S.}, {Davidsson, B.},
  {Schr\"oder, S. E.}, {Skorov, Y.}, {K\"uhrt, E.}, {Groussin, O.}, {Pajola,
  M.}, {Hviid, S. F.}, {Preusker, F.}, {Scholten, F.}, {A\'{}Hearn, M. F.},
  {Sierks, H.}, {Barbieri, C.}, {Lamy, P.}, {Rodrigo, R.}, {Koschny, D.},
  {Rickman, H.}, {Barucci, M. A.}, {Bertaux, J.-L.}, {Bertini, I.}, {Cremonese,
  G.}, {Da Deppo, V.}, {Debei, S.}, {De Cecco, M.}, {Fornasier, S.}, {Fulle,
  M.}, {Guti\'errez, P. J.}, {Ip, W.-H.}, {Jorda, L.}, {Knollenberg, J.},
  {Kramm, J. R.}, {K\"uppers, M.}, {Lara, L. M.}, {Lazzarin, M.}, {Lopez
  Moreno, J. J.}, {Marzari, F.}, {Michalik, H.}, {Naletto, G.}, {Sabau, L.},
  {Thomas, N.}, {Vincent, J.-B.}, {Wenzel, K.-P.}, {Agarwal, J.}, {G\"uttler,
  C.}, {Oklay, N.}, and {Tubiana, C.}]{Keller2015}
H.~U. Keller, {Mottola, S.}, {Davidsson, B.}, {Schr\"oder, S. E.}, {Skorov,
  Y.}, {K\"uhrt, E.}, {Groussin, O.}, {Pajola, M.}, {Hviid, S. F.}, {Preusker,
  F.}, {Scholten, F.}, {A\'{}Hearn, M. F.}, {Sierks, H.}, {Barbieri, C.},
  {Lamy, P.}, {Rodrigo, R.}, {Koschny, D.}, {Rickman, H.}, {Barucci, M. A.},
  {Bertaux, J.-L.}, {Bertini, I.}, {Cremonese, G.}, {Da Deppo, V.}, {Debei,
  S.}, {De Cecco, M.}, {Fornasier, S.}, {Fulle, M.}, {Guti\'errez, P. J.}, {Ip,
  W.-H.}, {Jorda, L.}, {Knollenberg, J.}, {Kramm, J. R.}, {K\"uppers, M.},
  {Lara, L. M.}, {Lazzarin, M.}, {Lopez Moreno, J. J.}, {Marzari, F.},
  {Michalik, H.}, {Naletto, G.}, {Sabau, L.}, {Thomas, N.}, {Vincent, J.-B.},
  {Wenzel, K.-P.}, {Agarwal, J.}, {G\"uttler, C.}, {Oklay, N.}, and {Tubiana,
  C.}
\newblock Insolation, erosion, and morphology of comet
  67p/churyumov-gerasimenko.
\newblock \emph{A\&A}, 583:\penalty0 A34, 2015.
\newblock \doi{10.1051/0004-6361/201525964}.
\newblock URL \url{https://doi.org/10.1051/0004-6361/201525964}.

\bibitem[{K{\"o}{\"o}p} et~al.(2018){K{\"o}{\"o}p}, {Heck}, {Busemann},
  {Davis}, {Greer}, {Maden}, {Meier}, and {Wieler}]{2018NatAs...2..709K}
L.~{K{\"o}{\"o}p}, P.~R. {Heck}, H.~{Busemann}, A.~M. {Davis}, J.~{Greer},
  C.~{Maden}, M.~M.~M. {Meier}, and R.~{Wieler}.
\newblock {High early solar activity inferred from helium and neon excesses in
  the oldest meteorite inclusions}.
\newblock \emph{Nature Astronomy}, 2:\penalty0 709--713, Jul 2018.
\newblock \doi{10.1038/s41550-018-0527-8}.

\bibitem[Kovaltsov et~al.(2014)Kovaltsov, Usoskin, Cliver, Dietrich, and
  Tylka]{kovaltsov2014fluence}
GA~Kovaltsov, IG~Usoskin, EW~Cliver, WF~Dietrich, and AJ~Tylka.
\newblock Fluence ordering of solar energetic proton events using cosmogenic
  radionuclide data.
\newblock \emph{Solar Physics}, 289\penalty0 (12):\penalty0 4691--4700, 2014.

\bibitem[Krimigis et~al.(2019)Krimigis, Decker, Roelof, Hill, Bostrom,
  Dialynas, Gloeckler, Hamilton, Keath, and Lanzerotti]{krimigis2019energetic}
Stamatios~M Krimigis, Robert~B Decker, Edmond~C Roelof, Matthew~E Hill, Carl~O
  Bostrom, Konstantinos Dialynas, George Gloeckler, Douglas~C Hamilton,
  Edward~P Keath, and Louis~J Lanzerotti.
\newblock Energetic charged particle measurements from voyager 2 at the
  heliopause and beyond.
\newblock \emph{Nature Astronomy}, 3\penalty0 (11):\penalty0 997--1006, 2019.

\bibitem[{K{\"u}hl} et~al.(2017){K{\"u}hl}, {Dresing}, {Heber}, and
  {Klassen}]{2017SoPh..292...10K}
P.~{K{\"u}hl}, N.~{Dresing}, B.~{Heber}, and A.~{Klassen}.
\newblock {Solar Energetic Particle Events with Protons Above 500 MeV Between
  1995 and 2015 Measured with SOHO/EPHIN}.
\newblock \emph{\solphys}, 292\penalty0 (1):\penalty0 10, Jan 2017.
\newblock \doi{10.1007/s11207-016-1033-8}.

\bibitem[L\"auter et~al.(2018)L\"auter, Kramer, Rubin, and Altwegg]{sty3103}
Matthias L\"auter, Tobias Kramer, Martin Rubin, and Kathrin Altwegg.
\newblock {Surface localization of gas sources on comet
  67P/Churyumov–Gerasimenko based on DFMS/COPS data}.
\newblock \emph{Monthly Notices of the Royal Astronomical Society},
  483\penalty0 (1):\penalty0 852--861, 11 2018.
\newblock ISSN 0035-8711.
\newblock \doi{10.1093/mnras/sty3103}.
\newblock URL \url{https://doi.org/10.1093/mnras/sty3103}.

\bibitem[Le~Roy et~al.(2015)Le~Roy, Altwegg, Balsiger, Berthelier, Bieler,
  Briois, Calmonte, Combi, De~Keyser, Dhooghe, et~al.]{LeRoy2015}
L{\'e}na Le~Roy, Kathrin Altwegg, Hans Balsiger, Jean-Jacques Berthelier, Andre
  Bieler, Christelle Briois, Ursina Calmonte, Michael~R Combi, Johan De~Keyser,
  Frederik Dhooghe, et~al.
\newblock Inventory of the volatiles on comet 67p/churyumov-gerasimenko from
  rosetta/rosina.
\newblock \emph{Astronomy \& Astrophysics}, 583:\penalty0 A1, 2015.

\bibitem[Lis et~al.(2019)Lis, Bockel\'ee-Morvan, G\"usten, Biver, Stutzki,
  Delorme, Dur\'an, Wiesemeyer, and Okada]{Dariusz2019}
Dariusz~C. Lis, Dominique Bockel\'ee-Morvan, Rolf G\"usten, Nicolas Biver,
  J\"urgen Stutzki, Yan Delorme, Carlos Dur\'an, Helmut Wiesemeyer, and Yoko
  Okada.
\newblock Terrestrial deuterium-to-hydrogen ratio in water in hyperactive
  comets.
\newblock \emph{A\&A}, 625:\penalty0 L5, 2019.
\newblock \doi{10.1051/0004-6361/201935554}.
\newblock URL \url{https://doi.org/10.1051/0004-6361/201935554}.

\bibitem[{Maquet}(2015)]{Maquet2015}
L.~{Maquet}.
\newblock {The recent dynamical history of comet 67P/Churyumov-Gerasimenko}.
\newblock \emph{\aap}, 579:\penalty0 A78, Jul 2015.
\newblock \doi{10.1051/0004-6361/201425461}.

\bibitem[{Massironi} et~al.(2015){Massironi}, {Simioni}, {Marzari},
  {Cremonese}, {Giacomini}, {Pajola}, {Jorda}, {Naletto}, {Lowry}, {El-Maarry},
  {Preusker}, {Scholten}, {Sierks}, {Barbieri}, {Lamy}, {Rodrigo}, {Koschny},
  {Rickman}, {Keller}, {A'Hearn}, {Agarwal}, {Auger}, {Barucci}, {Bertaux},
  {Bertini}, {Besse}, {Bodewits}, {Capanna}, {da Deppo}, {Davidsson}, {Debei},
  {de Cecco}, {Ferri}, {Fornasier}, {Fulle}, {Gaskell}, {Groussin},
  {Guti{\'e}rrez}, {G{\"u}ttler}, {Hviid}, {Ip}, {Knollenberg}, {Kovacs},
  {Kramm}, {K{\"u}hrt}, {K{\"u}ppers}, {La Forgia}, {Lara}, {Lazzarin}, {Lin},
  {Lopez Moreno}, {Magrin}, {Michalik}, {Mottola}, {Oklay}, {Pommerol},
  {Thomas}, {Tubiana}, and {Vincent}]{Massironi2015}
M.~{Massironi}, E.~{Simioni}, F.~{Marzari}, G.~{Cremonese}, L.~{Giacomini},
  M.~{Pajola}, L.~{Jorda}, G.~{Naletto}, S.~{Lowry}, M.~R. {El-Maarry},
  F.~{Preusker}, F.~{Scholten}, H.~{Sierks}, C.~{Barbieri}, P.~{Lamy},
  R.~{Rodrigo}, D.~{Koschny}, H.~{Rickman}, H.~U. {Keller}, M.~F. {A'Hearn},
  J.~{Agarwal}, A.-T. {Auger}, M.~A. {Barucci}, J.-L. {Bertaux}, I.~{Bertini},
  S.~{Besse}, D.~{Bodewits}, C.~{Capanna}, V.~{da Deppo}, B.~{Davidsson},
  S.~{Debei}, M.~{de Cecco}, F.~{Ferri}, S.~{Fornasier}, M.~{Fulle},
  R.~{Gaskell}, O.~{Groussin}, P.~J. {Guti{\'e}rrez}, C.~{G{\"u}ttler}, S.~F.
  {Hviid}, W.-H. {Ip}, J.~{Knollenberg}, G.~{Kovacs}, R.~{Kramm},
  E.~{K{\"u}hrt}, M.~{K{\"u}ppers}, F.~{La Forgia}, L.~M. {Lara},
  M.~{Lazzarin}, Z.-Y. {Lin}, J.~J. {Lopez Moreno}, S.~{Magrin}, H.~{Michalik},
  S.~{Mottola}, N.~{Oklay}, A.~{Pommerol}, N.~{Thomas}, C.~{Tubiana}, and J.-B.
  {Vincent}.
\newblock {Two independent and primitive envelopes of the bilobate nucleus of
  comet 67P}.
\newblock \emph{\nat}, 526:\penalty0 402--405, October 2015.
\newblock \doi{10.1038/nature15511}.

\bibitem[Matonti et~al.(2019)Matonti, Attree, Groussin, Jorda, Viseur, Hviid,
  Bouley, Nébouy, Auger, Lamy, Sierks, Naletto, Rodrigo, Koschny, Davidsson,
  Barucci, Bertaux, Bertini, Bodewits, Cremonese, Deppo, Debei, Cecco, Deller,
  Fornasier, Fulle, Gutiérrez, Güttler, Ip, Keller, Lara, Forgia, Lazzarin,
  Lucchetti, López-Moreno, Marzari, Massironi, Mottola, Oklay, Pajola, Penasa,
  Preusker, Rickman, Scholten, Shi, Toth, Tubiana, and
  Vincent]{matonti_bilobate_2019}
C.~Matonti, N.~Attree, O.~Groussin, L.~Jorda, S.~Viseur, S.~F. Hviid,
  S.~Bouley, D.~Nébouy, A.-T. Auger, P.~L. Lamy, H.~Sierks, G.~Naletto,
  R.~Rodrigo, D.~Koschny, B.~Davidsson, M.~A. Barucci, J.-L. Bertaux,
  I.~Bertini, D.~Bodewits, G.~Cremonese, V.~Da Deppo, S.~Debei, M.~De Cecco,
  J.~Deller, S.~Fornasier, M.~Fulle, P.~J. Gutiérrez, C.~Güttler, W.-H. Ip,
  H.~U. Keller, L.~M. Lara, F.~La Forgia, M.~Lazzarin, A.~Lucchetti, J.~J.
  López-Moreno, F.~Marzari, M.~Massironi, S.~Mottola, N.~Oklay, M.~Pajola,
  L.~Penasa, F.~Preusker, H.~Rickman, F.~Scholten, X.~Shi, I.~Toth, C.~Tubiana,
  and J.-B. Vincent.
\newblock Bilobate comet morphology and internal structure controlled by shear
  deformation.
\newblock \emph{Nature Geoscience}, page~1, February 2019.
\newblock ISSN 1752-0908.
\newblock \doi{10.1038/s41561-019-0307-9}.
\newblock URL \url{https://www.nature.com/articles/s41561-019-0307-9}.

\bibitem[Matthi{\"a} et~al.(2013)Matthi{\"a}, Berger, Mrigakshi, and
  Reitz]{matthia2013ready}
Daniel Matthi{\"a}, Thomas Berger, Alankrita~I Mrigakshi, and G{\"u}nther
  Reitz.
\newblock A ready-to-use galactic cosmic ray model.
\newblock \emph{Advances in Space Research}, 51\penalty0 (3):\penalty0
  329--338, 2013.

\bibitem[Mertens et~al.(2013)Mertens, Meier, Brown, Norman, and
  Xu]{mertens2013nairas}
Christopher~J Mertens, Matthias~M Meier, Steven Brown, Ryan~B Norman, and
  Xiaojing Xu.
\newblock Nairas aircraft radiation model development, dose climatology, and
  initial validation.
\newblock \emph{Space Weather}, 11\penalty0 (10):\penalty0 603--635, 2013.

\bibitem[{Morbidelli} and {Rickman}(2015)]{Morbidelli2015}
A.~{Morbidelli} and H.~{Rickman}.
\newblock {Comets as collisional fragments of a primordial planetesimal disk}.
\newblock \emph{\aap}, 583:\penalty0 A43, November 2015.
\newblock \doi{10.1051/0004-6361/201526116}.

\bibitem[Mousis et~al.(2017)Mousis, Drouard, Vernazza, Lunine, Monnereau,
  Maggiolo, Altwegg, Balsiger, Berthelier, Cessateur, De~Keyser, Fuselier,
  Gasc, Korth, Le~Deun, Mall, Marty, R{\`e}me, Rubin, Tzou, Waite, and
  P.~Wurz]{MousisEtAl2017}
O.~Mousis, A.~Drouard, P.~Vernazza, J.~I. Lunine, M.~Monnereau, R.~Maggiolo,
  K.~Altwegg, H.~Balsiger, J.-J. Berthelier, G.~Cessateur, J.~De~Keyser, S.~A.
  Fuselier, S.~Gasc, A.~Korth, T.~Le~Deun, U.~Mall, B.~Marty, H.~R{\`e}me,
  M.~Rubin, C.-Y. Tzou, J.~H. Waite, and P.~P.~Wurz.
\newblock {Impact of Radiogenic Heating on the Formation Conditions of Comet
  67P/Churyumov–Gerasimenko}.
\newblock \emph{Astrophys. J. Lett.}, 839\penalty0 (1):\penalty0 L4, apr 2017.
\newblock \doi{10.3847/2041-8213/aa6839}.

\bibitem[{Mousis} et~al.(2017){Mousis}, {Drouard}, {Vernazza}, {Lunine},
  {Monnereau}, {Maggiolo}, {Altwegg}, {Balsiger}, {Berthelier}, {Cessateur},
  {De Keyser}, {Fuselier}, {Gasc}, {Korth}, {Le Deun}, {Mall}, {Marty},
  {R{\`e}me}, {Rubin}, {Tzou}, {Waite}, and {Wurz}]{Mousis2017}
O.~{Mousis}, A.~{Drouard}, P.~{Vernazza}, J.~I. {Lunine}, M.~{Monnereau},
  R.~{Maggiolo}, K.~{Altwegg}, H.~{Balsiger}, J.-J. {Berthelier},
  G.~{Cessateur}, J.~{De Keyser}, S.~A. {Fuselier}, S.~{Gasc}, A.~{Korth},
  T.~{Le Deun}, U.~{Mall}, B.~{Marty}, H.~{R{\`e}me}, M.~{Rubin}, C.-Y. {Tzou},
  J.~H. {Waite}, and P.~{Wurz}.
\newblock {Impact of Radiogenic Heating on the Formation Conditions of Comet
  67P/Churyumov-Gerasimenko}.
\newblock \emph{\apjl}, 839:\penalty0 L4, April 2017.
\newblock \doi{10.3847/2041-8213/aa6839}.

\bibitem[{Mumma} et~al.(1993){Mumma}, {Weissman}, and {Stern}]{Mumma1993}
M.~J. {Mumma}, P.~R. {Weissman}, and S.~A. {Stern}.
\newblock {Comets and the origin of the solar system - Reading the Rosetta
  Stone}.
\newblock In E.~H. {Levy} and J.~I. {Lunine}, editors, \emph{Protostars and
  Planets III}, pages 1177--1252, 1993.

\bibitem[Norman et~al.(2014)Norman, Gronoff, and
  Mertens]{norman_influence_2014}
Ryan~B. Norman, Guillaume Gronoff, and Christopher~J. Mertens.
\newblock Influence of dust loading on atmospheric ionizing radiation on mars.
\newblock \emph{\jgr: Space Physics}, 2014.
\newblock ISSN 2169-9402.
\newblock \doi{10.1002/2013JA019351}.
\newblock URL
  \url{http://onlinelibrary.wiley.com/doi/10.1002/2013JA019351/abstract}.

\bibitem[O'Neill et~al.(2014)O'Neill, Golge, and Slaba]{o2014badhwar}
PM~O'Neill, S~Golge, and TC~Slaba.
\newblock Badhwar-o'neill 2014 galactic cosmic ray flux model description.
\newblock 2014.

\bibitem[P{\"a}tzold et~al.(2016)P{\"a}tzold, Andert, Hahn, Asmar, Barriot,
  Bird, H{\"a}usler, Peter, Tellmann, Gr{\"u}n, et~al.]{Paetzold2016}
M~P{\"a}tzold, T~Andert, M~Hahn, SW~Asmar, J-P Barriot, MK~Bird, B~H{\"a}usler,
  K~Peter, S~Tellmann, Eberhard Gr{\"u}n, et~al.
\newblock A homogeneous nucleus for comet 67p/churyumov--gerasimenko from its
  gravity field.
\newblock \emph{Nature}, 530\penalty0 (7588):\penalty0 63, 2016.

\bibitem[{P{\"a}tzold} et~al.(2019){P{\"a}tzold}, {Andert}, {Hahn}, {Barriot},
  {Asmar}, {H{\"a}usler}, {Bird}, {Tellmann}, {Oschlisniok}, and
  {Peter}]{Patzold2019}
M.~{P{\"a}tzold}, T.~P. {Andert}, M.~{Hahn}, J.-P. {Barriot}, S.~W. {Asmar},
  B.~{H{\"a}usler}, M.~K. {Bird}, S.~{Tellmann}, J.~{Oschlisniok}, and
  K.~{Peter}.
\newblock {The Nucleus of comet 67P/Churyumov-Gerasimenko - Part I: The global
  view - nucleus mass, mass-loss, porosity, and implications}.
\newblock \emph{\mnras}, 483:\penalty0 2337--2346, February 2019.
\newblock \doi{10.1093/mnras/sty3171}.

\bibitem[Pavlov et~al.(2014)Pavlov, Pavlov, Ostryakov, Vasilyev, Mahaffy, and
  Steele]{Pavlov2014}
A.~A. Pavlov, A.~K. Pavlov, V.~M. Ostryakov, G.~I. Vasilyev, P.~Mahaffy, and
  A.~Steele.
\newblock Alteration of the carbon and nitrogen isotopic composition in the
  martian surface rocks due to cosmic ray exposure.
\newblock \emph{\jgr: Planets}, 119\penalty0 (6):\penalty0 1390--1402, 2014.
\newblock ISSN 2169-9100.
\newblock \doi{10.1002/2014JE004615}.
\newblock URL \url{http://dx.doi.org/10.1002/2014JE004615}.

\bibitem[{Poluianov} et~al.(2018){Poluianov}, {Kovaltsov}, and
  {Usoskin}]{2018A&A...618A..96P}
S.~{Poluianov}, G.~A. {Kovaltsov}, and I.~G. {Usoskin}.
\newblock {Solar energetic particles and galactic cosmic rays over millions of
  years as inferred from data on cosmogenic $^{26}$Al in lunar samples}.
\newblock \emph{\aap}, 618:\penalty0 A96, Oct 2018.
\newblock \doi{10.1051/0004-6361/201833561}.

\bibitem[{Portegies Zwart} et~al.(2018){Portegies Zwart}, {Pelupessy}, {van
  Elteren}, {Wijnen}, and {Lugaro}]{2018A&A...616A..85P}
S.~{Portegies Zwart}, I.~{Pelupessy}, A.~{van Elteren}, T.~P.~G. {Wijnen}, and
  M.~{Lugaro}.
\newblock {The consequences of a nearby supernova on the early solar system}.
\newblock \emph{\aap}, 616:\penalty0 A85, Aug 2018.
\newblock \doi{10.1051/0004-6361/201732060}.

\bibitem[Quirico et~al.(2016)Quirico, Moroz, Schmitt, Arnold, Faure, Beck,
  Bonal, Ciarniello, Capaccioni, Filacchione, Erard, Leyrat, Bockelée-Morvan,
  Zinzi, Palomba, Drossart, Tosi, Capria, Sanctis, Raponi, Fonti, Mancarella,
  Orofino, Barucci, Blecka, Carlson, Despan, Faure, Fornasier, Gudipati,
  Longobardo, Markus, Mennella, Merlin, Piccioni, Rousseau, and
  Taylor]{QUIRICO201632}
E.~Quirico, L.V. Moroz, B.~Schmitt, G.~Arnold, M.~Faure, P.~Beck, L.~Bonal,
  M.~Ciarniello, F.~Capaccioni, G.~Filacchione, S.~Erard, C.~Leyrat,
  D.~Bockelée-Morvan, A.~Zinzi, E.~Palomba, P.~Drossart, F.~Tosi, M.T. Capria,
  M.C.~De Sanctis, A.~Raponi, S.~Fonti, F.~Mancarella, V.~Orofino, A.~Barucci,
  M.I. Blecka, R.~Carlson, D.~Despan, A.~Faure, S.~Fornasier, M.S. Gudipati,
  A.~Longobardo, K.~Markus, V.~Mennella, F.~Merlin, G.~Piccioni, B.~Rousseau,
  and F.~Taylor.
\newblock Refractory and semi-volatile organics at the surface of comet
  67p/churyumov-gerasimenko: Insights from the virtis/rosetta imaging
  spectrometer.
\newblock \emph{Icarus}, 272:\penalty0 32 -- 47, 2016.
\newblock ISSN 0019-1035.
\newblock \doi{https://doi.org/10.1016/j.icarus.2016.02.028}.
\newblock URL
  \url{http://www.sciencedirect.com/science/article/pii/S001910351600097X}.

\bibitem[{Rickman} et~al.(2015){Rickman}, {Marchi}, {A'Hearn}, {Barbieri},
  {El-Maarry}, {G{\"u}ttler}, {Ip}, {Keller}, {Lamy}, {Marzari}, {Massironi},
  {Naletto}, {Pajola}, {Sierks}, {Koschny}, {Rodrigo}, {Barucci}, {Bertaux},
  {Bertini}, {Cremonese}, {Da Deppo}, {Debei}, {De Cecco}, {Fornasier},
  {Fulle}, {Groussin}, {Guti{\'e}rrez}, {Hviid}, {Jorda}, {Knollenberg},
  {Kramm}, {K{\"u}hrt}, {K{\"u}ppers}, {Lara}, {Lazzarin}, {Lopez Moreno},
  {Michalik}, {Sabau}, {Thomas}, {Vincent}, and {Wenzel}]{Rickman2015}
H.~{Rickman}, S.~{Marchi}, M.~F. {A'Hearn}, C.~{Barbieri}, M.~R. {El-Maarry},
  C.~{G{\"u}ttler}, W.-H. {Ip}, H.~U. {Keller}, P.~{Lamy}, F.~{Marzari},
  M.~{Massironi}, G.~{Naletto}, M.~{Pajola}, H.~{Sierks}, D.~{Koschny},
  R.~{Rodrigo}, M.~A. {Barucci}, J.-L. {Bertaux}, I.~{Bertini}, G.~{Cremonese},
  V.~{Da Deppo}, S.~{Debei}, M.~{De Cecco}, S.~{Fornasier}, M.~{Fulle},
  O.~{Groussin}, P.~J. {Guti{\'e}rrez}, S.~F. {Hviid}, L.~{Jorda},
  J.~{Knollenberg}, J.-R. {Kramm}, E.~{K{\"u}hrt}, M.~{K{\"u}ppers}, L.~M.
  {Lara}, M.~{Lazzarin}, J.~J. {Lopez Moreno}, H.~{Michalik}, L.~{Sabau},
  N.~{Thomas}, J.-B. {Vincent}, and K.-P. {Wenzel}.
\newblock {Comet 67P/Churyumov-Gerasimenko: Constraints on its origin from
  OSIRIS observations}.
\newblock \emph{\aap}, 583:\penalty0 A44, November 2015.
\newblock \doi{10.1051/0004-6361/201526093}.

\bibitem[Rodr{\'i}guez-Gas{\'e}n et~al.(2014)Rodr{\'i}guez-Gas{\'e}n, Aran,
  Sanahuja, Jacobs, and Poedts]{RodriguezGasen2014}
R.~Rodr{\'i}guez-Gas{\'e}n, A.~Aran, B.~Sanahuja, C.~Jacobs, and S.~Poedts.
\newblock Variation of proton flux profiles with the observer's latitude in
  simulated gradual sep events.
\newblock \emph{Solar Physics}, 289\penalty0 (5):\penalty0 1745--1762, May
  2014.
\newblock ISSN 1573-093X.
\newblock \doi{10.1007/s11207-013-0442-1}.
\newblock URL \url{https://doi.org/10.1007/s11207-013-0442-1}.

\bibitem[Rotundi et~al.(2015)Rotundi, Sierks, Della~Corte, Fulle, Gutierrez,
  Lara, Barbieri, Lamy, Rodrigo, Koschny, Rickman, Keller, L{\'o}pez-Moreno,
  Accolla, Agarwal, A{\textquoteright}Hearn, Altobelli, Angrilli, Barucci,
  Bertaux, Bertini, Bodewits, Bussoletti, Colangeli, Cosi, Cremonese, Crifo,
  Da~Deppo, Davidsson, Debei, De~Cecco, Esposito, Ferrari, Fornasier, Giovane,
  Gustafson, Green, Groussin, Gr{\"u}n, G{\"u}ttler, Herranz, Hviid, Ip,
  Ivanovski, Jer{\'o}nimo, Jorda, Knollenberg, Kramm, K{\"u}hrt, K{\"u}ppers,
  Lazzarin, Leese, L{\'o}pez-Jim{\'e}nez, Lucarelli, Lowry, Marzari, Epifani,
  McDonnell, Mennella, Michalik, Molina, Morales, Moreno, Mottola, Naletto,
  Oklay, Ortiz, Palomba, Palumbo, Perrin, Rodr{\'\i}guez, Sabau, Snodgrass,
  Sordini, Thomas, Tubiana, Vincent, Weissman, Wenzel, Zakharov, and
  Zarnecki]{Rotundiaaa3905}
Alessandra Rotundi, Holger Sierks, Vincenzo Della~Corte, Marco Fulle, Pedro~J.
  Gutierrez, Luisa Lara, Cesare Barbieri, Philippe~L. Lamy, Rafael Rodrigo,
  Detlef Koschny, Hans Rickman, Horst~Uwe Keller, Jos{\'e}~J. L{\'o}pez-Moreno,
  Mario Accolla, Jessica Agarwal, Michael~F. A{\textquoteright}Hearn, Nicolas
  Altobelli, Francesco Angrilli, M.~Antonietta Barucci, Jean-Loup Bertaux,
  Ivano Bertini, Dennis Bodewits, Ezio Bussoletti, Luigi Colangeli, Massimo
  Cosi, Gabriele Cremonese, Jean-Francois Crifo, Vania Da~Deppo, Bj{\"o}rn
  Davidsson, Stefano Debei, Mariolino De~Cecco, Francesca Esposito, Marco
  Ferrari, Sonia Fornasier, Frank Giovane, Bo~Gustafson, Simon~F. Green,
  Olivier Groussin, Eberhard Gr{\"u}n, Carsten G{\"u}ttler, Miguel~L. Herranz,
  Stubbe~F. Hviid, Wing Ip, Stavro Ivanovski, Jos{\'e}~M. Jer{\'o}nimo, Laurent
  Jorda, Joerg Knollenberg, Rainer Kramm, Ekkehard K{\"u}hrt, Michael
  K{\"u}ppers, Monica Lazzarin, Mark~R. Leese, Antonio~C.
  L{\'o}pez-Jim{\'e}nez, Francesca Lucarelli, Stephen~C. Lowry, Francesco
  Marzari, Elena~Mazzotta Epifani, J.~Anthony~M. McDonnell, Vito Mennella,
  Harald Michalik, Antonio Molina, Rafael Morales, Fernando Moreno, Stefano
  Mottola, Giampiero Naletto, Nilda Oklay, Jos{\'e}~L. Ortiz, Ernesto Palomba,
  Pasquale Palumbo, Jean-Marie Perrin, Julio Rodr{\'\i}guez, Lola Sabau, Colin
  Snodgrass, Roberto Sordini, Nicolas Thomas, Cecilia Tubiana, Jean-Baptiste
  Vincent, Paul Weissman, Klaus-Peter Wenzel, Vladimir Zakharov, and John~C.
  Zarnecki.
\newblock Dust measurements in the coma of comet 67p/churyumov-gerasimenko
  inbound to the sun.
\newblock \emph{Science}, 347\penalty0 (6220), 2015.
\newblock ISSN 0036-8075.
\newblock \doi{10.1126/science.aaa3905}.
\newblock URL \url{https://science.sciencemag.org/content/347/6220/aaa3905}.

\bibitem[Rubin et~al.(2015)Rubin, Altwegg, Balsiger, Bar-Nun, Berthelier,
  Bieler, Bochsler, Briois, Calmonte, Combi, et~al.]{Rubin2015}
Martin Rubin, Kathrin Altwegg, Hans Balsiger, A~Bar-Nun, J-J Berthelier,
  Andr{\'e} Bieler, Peter Bochsler, Christelle Briois, U~Calmonte, M~Combi,
  et~al.
\newblock Molecular nitrogen in comet 67p/churyumov-gerasimenko indicates a low
  formation temperature.
\newblock \emph{Science}, 348\penalty0 (6231):\penalty0 232--235, 2015.

\bibitem[Saxena et~al.(2019)Saxena, Killen, Airapetian, Petro, Curran, and
  Mandell]{Saxena_2019}
Prabal Saxena, Rosemary~M. Killen, Vladimir Airapetian, Noah~E. Petro,
  Natalie~M. Curran, and Avi~M. Mandell.
\newblock Was the sun a slow rotator? sodium and potassium constraints from the
  lunar regolith.
\newblock \emph{The Astrophysical Journal}, 876\penalty0 (1):\penalty0 L16, may
  2019.
\newblock \doi{10.3847/2041-8213/ab18fb}.
\newblock URL \url{https://doi.org/10.3847\%2F2041-8213\%2Fab18fb}.

\bibitem[Schwartz et~al.(2018)Schwartz, Michel, Jutzi, Marchi, Zhang, and
  Richardson]{schwartz2018catastrophic}
Stephen~R Schwartz, Patrick Michel, Martin Jutzi, Simone Marchi, Yun Zhang, and
  Derek~C Richardson.
\newblock Catastrophic disruptions as the origin of bilobate comets.
\newblock \emph{Nature astronomy}, 2\penalty0 (5):\penalty0 379, 2018.

\bibitem[Sheel et~al.(2012)Sheel, Haider, Withers, Kozarev, Jun, Kang, Gronoff,
  and Wedlund]{sheel_numerical_2012}
Varun Sheel, S.~A. Haider, Paul Withers, K.~Kozarev, I.~Jun, S.~Kang,
  G.~Gronoff, and C.~Simon Wedlund.
\newblock Numerical simulation of the effects of a solar energetic particle
  event on the ionosphere of mars.
\newblock \emph{J. Geophys. Res.}, 117:\penalty0 13 PP., May 2012.
\newblock \doi{201210.1029/2011JA017455}.
\newblock URL \url{http://www.agu.org/pubs/crossref/2012/2011JA017455.shtml}.

\bibitem[Simnett(2017)]{simnett_anomalous_2017}
George~M. Simnett.
\newblock The {Anomalous} {Cosmic} {Rays}.
\newblock In \emph{Energetic {Particles} in the {Heliosphere}}, number 438 in
  Astrophysics and {Space} {Science} {Library}, pages 189--200. Springer
  International Publishing, 2017.
\newblock ISBN 978-3-319-43493-3 978-3-319-43495-7.

\bibitem[Smart et~al.(2006)Smart, Shea, and McCracken]{Smart2006}
DF~Smart, MA~Shea, and KG~McCracken.
\newblock The carrington event: Possible solar proton intensity--time profile.
\newblock \emph{Advances in Space Research}, 38\penalty0 (2):\penalty0
  215--225, 2006.

\bibitem[{SPENVIS}(2019)]{SPENVIS2019}
{SPENVIS}.
\newblock - {Space} {Environment}, {Effects}, and {Education} {System}, 2019.
\newblock URL \url{https://www.spenvis.oma.be/}.

\bibitem[{Stern}(1988)]{Stern1988Icar}
S.~A. {Stern}.
\newblock {Collisions in the Oort Cloud}.
\newblock \emph{\icarus}, 73:\penalty0 499--507, March 1988.
\newblock \doi{10.1016/0019-1035(88)90059-0}.

\bibitem[{Stern}(2003)]{Stern2003}
S.~A. {Stern}.
\newblock {The evolution of comets in the Oort cloud and Kuiper belt}.
\newblock \emph{\nat}, 424:\penalty0 639--642, August 2003.

\bibitem[{Stern} and {Shull}(1988)]{Stern1988Natur}
S.~A. {Stern} and J.~M. {Shull}.
\newblock {The influence of supernovae and passing stars on comets in the Oort
  cloud}.
\newblock \emph{\nat}, 332:\penalty0 407--411, March 1988.
\newblock \doi{10.1038/332407a0}.

\bibitem[{Stern} and {Weissman}(2001)]{Stern2001}
S.~A. {Stern} and P.~R. {Weissman}.
\newblock {Rapid collisional evolution of comets during the formation of the
  Oort cloud}.
\newblock \emph{\nat}, 409:\penalty0 589--591, February 2001.

\bibitem[{Torres} et~al.(2012){Torres}, {Cillis}, {Lacki}, and
  {Rephaeli}]{Torres2012}
D.~F. {Torres}, A.~{Cillis}, B.~{Lacki}, and Y.~{Rephaeli}.
\newblock Building up the spectrum of cosmic rays in star-forming regions.
\newblock \emph{\mnras}, 423:\penalty0 822--830, June 2012.
\newblock \doi{10.1111/j.1365-2966.2012.20920.x}.

\bibitem[Velinov and Mateev(2008)]{Velinov2008}
P.~I.~Y. Velinov and L.~Mateev.
\newblock {Analytical approach to cosmic ray ionization by nuclei with charge Z
  in the middle atmosphere Distribution of galactic CR effects}.
\newblock \emph{Adv. Sp. Res.}, 42:\penalty0 1586--1592, 2008.
\newblock \doi{10.1016/j.asr.2007.12.008}.

\bibitem[Wolff et~al.(2012)Wolff, Bigler, Curran, Dibb, Frey, Legrand, and
  McConnell]{Wolff2012}
E.~W. Wolff, M.~Bigler, M.~A.~J. Curran, J.~E. Dibb, M.~M. Frey, M.~Legrand,
  and J.~R. McConnell.
\newblock The carrington event not observed in most ice core nitrate records.
\newblock \emph{Geophysical Research Letters}, 39\penalty0 (8), 2012.
\newblock \doi{10.1029/2012GL051603}.
\newblock URL
  \url{https://agupubs.onlinelibrary.wiley.com/doi/abs/10.1029/2012GL051603}.

\bibitem[{Wurz}(2005)]{2005ESASP.600E..44W}
P.~{Wurz}.
\newblock {Solarwind Composition}.
\newblock In \emph{The Dynamic Sun: Challenges for Theory and Observations},
  volume 600 of \emph{ESA Special Publication}, page 44.1, December 2005.

\end{thebibliography}

\clearpage



\end{document}